\documentclass[fleqn,usenatbib]{mnras}

\usepackage{newtxtext,newtxmath}
\usepackage{comment}
\usepackage{longtable}
\usepackage{listings}
\usepackage[T1]{fontenc}
\usepackage{xcolor, color}
\usepackage{float}

\DeclareRobustCommand{\VAN}[3]{#2}
\let\VANthebibliography\thebibliography
\def\thebibliography{\DeclareRobustCommand{\VAN}[3]{##3}\VANthebibliography}
\defcitealias{VargasMelatos2023}{VM23}
\defcitealias{VargasMelatos2024}{VM24}
\defcitealias{VargasMelatos2025}{VM25}

\usepackage{graphicx}	
\usepackage{amsmath}	
\usepackage{bm}
\usepackage{physics}
\usepackage{tabularx}
\usepackage{parskip}
\newcommand{\tempoDOS}{$\mathrm{{\scriptstyle TEMPO2}}$}

\newcommand{\temponest}{$\mathrm{{\scriptstyle TEMPONEST}}$}

\title[Bayesian torque estimation for P574 pulsars]{Hierarchical Bayesian estimation of population-level torque law parameters from $68$ young radio pulsars observed with the Murriyang telescope}

\author[A.~F.~Vargas et al.]{\parbox{\linewidth}{\flushleft
Andr\'es F. Vargas$^{1,2}$\thanks{E-mail: a.vargas@unimelb.edu.au},
Andrew Melatos$^{1,2}$,
Julian B. Carlin$^{1,2,3}$, 
Marcus E. Lower$^{4,5}$,
Simon Johnston$^{5}$,
and Patrick Weltevrede$^{6}$.
}\\\\ 
$^{1}$School of Physics, University of Melbourne, Parkville, VIC 3010, Australia\\
$^{2}$OzGrav: The Australian Research Council Centre of Excellence for Gravitational-wave Discovery, University of Melbourne, Parkville, VIC 3010, Australia\\
$^{3}$Department of Infectious Diseases, University of Melbourne, Parkville, VIC 3010, Australia\\
$^{4}$Centre for Astrophysics and Supercomputing, Swinburne University of Technology, PO Box 218, Hawthorn, VIC 3122, Australia\\
$^{5}$Australia Telescope National Facility, CSIRO, Space and Astronomy, PO Box 76, Epping, NSW 1710, Australia\\
$^{6}$Jodrell Bank Centre for Astrophysics, The University of Manchester, Alan Turing Building, Manchester, M13 9PL, United Kingdom
}

\date{Accepted XXX. Received YYY; in original form ZZZ}

\pubyear{2024}

\begin{document}
\label{firstpage}
\pagerange{\pageref{firstpage}--\pageref{lastpage}}
\maketitle

\begin{abstract}

\noindent The measured braking index, $n=\nu \ddot{\nu}/\dot{\nu}^2$, of a rotation-powered pulsar with spin frequency $\nu$ (where an overdot symbolizes a time derivative) and braking torque $K \nu^{n_{\rm pl}}$, features secular and stochastic anomalies arising from $\dot{K} \neq 0$ and random torque noise respectively. Previous studies quantified the variance $\langle n^{2} \rangle = (n_{\rm pl}+\dot{K}_{\rm dim})^{2}+\sigma_{\rm dim}^{2}$, where the secular anomaly, $\dot{K}_{\rm dim}$, is inversely proportional to the characteristic time-scale $\tau_{K}$ over which $K$ varies; the stochastic anomaly, $\sigma_{\rm dim}^{2} = \sigma_{\ddot{\nu}}^{2}\nu^{2}\gamma_{\ddot{\nu}}^{-2}\dot{\nu}^{-4}T_{\rm obs}^{-1}$, is a function of the timing noise amplitude $\sigma_{\ddot{\nu}}$, a damping time-scale $\gamma_{\ddot{\nu}}^{-1}$ and the total observing time $T_{\rm obs}$; and the average is taken over an ensemble of random realizations of the noise process. Here, we use a hierarchical Bayesian scheme, based on the formula for $\langle n^{2} \rangle$, to infer the population-level distribution of $n_{\rm pl}+\dot{K}_{\rm dim}$ for a sample of $68$ young radio pulsars, observed for $\gtrsim 10~{\rm years}$ with Murriyang, the 64-m Parkes radio telescope. Upon  assuming that the $n_{\rm pl}+\dot{K}_{\rm dim}$ values are drawn from a population-level Gaussian, ${\cal N}(\mu_{\rm pl}, \sigma_{\rm pl})$, the Bayesian scheme returns the mean $\mu_{\rm pl} = 9.95^{+5.58}_{-5.26}$ and standard deviation $\sigma_{\rm pl}=10.89^{+5.14}_{-3.69}$. At a per-pulsar level it returns posterior medians satisfying $-13.86 \leq n_{\rm pl}+\dot{K}_{\rm dim} \leq 30.38$. The secular anomaly dominates the stochastic anomaly, with posterior medians satisfying $|n_{\rm pl} + \dot{K}_{\rm dim}| \geq \sigma_{\rm dim}$ in 10 out of 68 objects. The inference results imply that some mechanism other than electromagnetic or gravitational radiation reaction with $\dot{K} = 0$ operates in at least 66 out of 68 analyzed pulsars, in line with previous observational studies. The results are also consistent with $\tau_K \lesssim \nu /\vert\dot{\nu}\vert$, with implications for the physical mechanism causing $\dot{K} \neq 0$.\\
\end{abstract}

\begin{keywords}
methods: data analysis -- pulsars: general -- stars: rotation
\end{keywords}



\section{Introduction}
\label{Sec:Introduction}

The long-term evolution of the braking torque of a rotation-powered pulsar probes the pulsar's magnetosphere and interior~\citep{BlandfordRomani1988}. It is studied through timing experiments by measuring the braking index,

\begin{equation}
    n=\frac{\nu \ddot{\nu}}{\dot{\nu}^{2}},
    \label{Eq_SecI:Intro_n_def}
\end{equation}

where $\nu$ is the pulsar's spin frequency, and the overdot symbolizes a derivative with respect to time. Plausible physical theories suggest the braking torque follows a power law $\dot{\nu}= K\nu^{n_{\rm pl}}$, with $K$ constant and $n=n_{\rm pl}$ in the absence of stochastic fluctuations due to instrumental factors, interstellar propagation or intrinsic (achromatic) rotational jitter. Examples of power-law braking include electromagnetic radiation reaction, with $n_{\rm pl} \approx 2l-1$, where $l$ is the leading multipole order~\citep{GunnOstriker1969,Goldreich1970,Melatos1997,BucciantiniThompson2006,ContopoulosSpitkovsky2006,KouTong2015,Petri2015,Petri2017,AraujoDeLorenci2024}, and gravitational radiation reaction, with $5\leq n_{\rm pl} \leq 7$ for mass and current quadrupole emission~\citep{PapaloizouPringle1978,Thorne1980,Andersson1998,OwenLindblom1998}.

Pulsar timing experiments typically measure braking indices in the range $ 2\lesssim n\leq3$, consistent with electromagnetic braking~\citep{EspinozaLyne2017}, for pulsars that do not involve correcting for major rotational glitches --- which are impulsive spin-up events --- e.g. PSR J0534$+$2200 with $n=2.51\pm0.01$~\citep{LyneAG1993}, and PSR J1640$-$4631 with $n=3.15\pm0.03$~\citep{ArchibaldGotthelf2016}. Additionally, some pulsars for which it is possible to correct for post-glitch recoveries, return measured braking indices consistent with electromagnetic braking~\citep{LynePritchard1996,EspinozaLyne2017}, e.g. J0537$-$6910 with $n=2.7\pm0.4$~\citep{AkbalGugercinoglu2021}. However, timing experiments also measure `anomalous' braking indices, i.e., high-$\vert n \vert$ values satisfying $3 \ll \vert n \vert \lesssim 10^{6}$~\citep{JohnstonGalloway1999,ChukwudeChidiOdo2016, LowerBailes2020, ParthasarathyJohnston2020, OnuchukwuLegahara2024}, especially in pulsars that show high glitch activity and timing noise~\citep{Cordes1980,ArzoumanianNice1994,JohnstonGalloway1999,UramaLink2006,LowerJohnston2021}. 

In this paper, we consider two phenomenological modifications of $
\dot{\nu} = K\nu^{n_{\rm pl}}$ to explain anomalous braking indices: (i) $K$ evolves secularly $(\dot{K} \neq 0)$, and (ii) $K\nu^{n_{\rm pl}}$ is supplemented by a stochastic torque, which dominates $\ddot{\nu}$ over typical observational time-scales. Importantly, scenarios (i) and (ii) may coexist with each other and other plausible explanations~\citep{ChukwudeBaiden2010,ColesHobbs2011,OnuchukwuLegahara2024}. In scenario (i), $K$ increases (or decreases) secularly on a time-scale $\tau_{K}$ (much) shorter than the pulsar's spin-down time-scale $\tau_{\rm sd}=\nu/(2\vert \dot{\nu} \vert)$, implying $\vert n \vert \approx \vert n_{\rm pl}+(\dot{K}/K)(\nu/\dot{\nu}) \vert \approx \tau_{\rm sd}/\tau_{K} \gg n_{\rm pl}$. Examples of (i) include (counter-)alignment of the rotation and magnetic axes~\citep{Goldreich1970,LinkEpstein1997,Melatos2000,BarsukovPolyakova2009,JohnstonKarastergiou2017,AbolmasovBiryukov2024}, magnetic field evolution~\citep{TaurisKonar2001,PonsVigano2012}, Hall drift~\citep{BransgroveLevin2025}, precession~\citep{Melatos2000,BarsukovTsygan2010,WassermanCordes2022}, or magnetospheric switching~\citep{LyneHobbs2010,StairsLyne2019}. In scenario (ii), one adds a stochastic torque $\xi(t)$ to $K\nu(t)^{n_{\rm pl}}$. For example, $\xi(t)$ may arise due to crust-superfluid relaxation processes~\citep{SedrakianCordes1998,AlparBaykal2006,GugercinogluAlpar2014,Gugercinoglu2017,LowerJohnston2021}, or achromatic timing noise inherent to the star's interior or crust~\citep{CordesDowns1985,Jones1990,MelatosLink2014,ChukwudeChidiOdo2016}.

\cite{VargasMelatos2024} (henceforth~\citetalias{VargasMelatos2024}) combined scenarios (i) and (ii) in the context of an idealized, phenomenological model in which the fluctuations $\delta\ddot{\nu}(t) = \ddot{\nu}(t)-d[K\nu(t)^{n_{\rm pl}}]/dt$, with $\dot{K} \neq 0$, follow mean-reverting (damped) Brownian motion with $d[\delta \ddot{\nu}(t)]/dt = -\gamma_{\ddot{\nu}}\delta\ddot{\nu}(t)+\xi(t)$, where $\gamma_{\ddot{\nu}}^{-1}$ is the mean-reversion time-scale, and $\xi(t)$ is a noisy driver~[\citep{VargasMelatos2023}; henceforth~\citetalias{VargasMelatos2023}].  With these assumptions, \citetalias{VargasMelatos2024} derived analytically and verified through Monte Carlo simulations that the variance of the measured $n$ satisfies the predictive, falsifiable formula 

\begin{equation}
    \langle n^{2} \rangle = (n_{\rm pl}+\dot{K}_{\rm dim})^{2}+\frac{\sigma^{2}_{\ddot{\nu}}\nu^{2}}{\gamma_{\ddot{\nu}}^{2}\dot{\nu}^{4}T_{\rm obs}},
    \label{Eq_SecI:Variance_n}
\end{equation}

where the angular brackets denote an ensemble average over random realizations of $\xi(t)$, $\sigma^{2}_{\ddot{\nu}} \propto \langle \xi(t)\xi(t')\rangle$ is the squared timing noise amplitude, $T_{\rm obs}$ is the total observing time, and one has $\dot{K}_{\dim} = (\dot{K}/K)(\nu/\dot{\nu}) \propto \tau_{\rm sd}/\tau_{K}$. Equation~(\ref{Eq_SecI:Variance_n}) implies that anomalous braking indices, i.e. $\langle n \rangle \gg 1$, occur for $K_{\rm dim} \gg 1$ [scenario (i)], and/or $\sigma^{2}_{\ddot{\nu}}\nu^{2}/\gamma_{\ddot{\nu}}^{2}\dot{\nu}^{4}T_{\rm obs} \gg 1$ [scenario (ii)]. 

Recently, \cite{VargasMelatos2025} (henceforth~\citetalias{VargasMelatos2025}) combined equation~(\ref{Eq_SecI:Variance_n}) with a hierarchical Bayesian scheme and showed how to infer the distributions of $n_{\rm pl}$ and $\sigma^{2}_{\ddot{\nu}}/\gamma_{\ddot{\nu}}^{2}$ for a synthetic pulsar population of size $M \geq 50$, given per-pulsar measurements of the nominal braking index $n^{(m)}_{\rm meas}$ (which can be anomalous or not), the root-mean-square of the timing residuals $S_{\rm meas}^{(m)}$, and their associated uncertainties $\Delta n^{(m)}_{\rm meas}$ and $\Delta S_{\rm meas}^{(m)}$, respectively. Monte Carlo simulations show that for $M \geq 50$, the scheme infers per-pulsar posteriors which contain $\gtrsim 70\%$ of the injected $n_{\rm pl}$ and $\sigma^{2}_{\ddot{\nu}}/\gamma_{\ddot{\nu}}^{2}$ values within their $90\%$ credible intervals. 

In this paper we use the hierarchical Bayesian scheme developed by~\citetalias{VargasMelatos2025} to infer the distribution of the quantities $n_{\rm pl}$ and $\sigma^{2}_{\ddot{\nu}}/\gamma_{\ddot{\nu}}^{2}$ for a sample of $68$ young radio pulsars, observed for $\gtrsim 10$ years with Murriyang, the CSIRO Parkes 64-m radio telescope. The sample was analyzed previously by~\cite{ParthasarathyShannon2019} and~\cite{ParthasarathyJohnston2020}. The paper is structured as follows. In Section~\ref{Sec:Methods}, we introduce the Brownian rotational model from~\citetalias{VargasMelatos2024} and the associated hierarchical Bayesian scheme from~\citetalias{VargasMelatos2025}. In Section~\ref{Sec:Observations}, we briefly describe the observing program and the young pulsar sample. Section~\ref{Sec:Analysis} uses the hierarchical Bayesian scheme to infer the per-pulsar and population-level posterior distributions for $n_{\rm pl}+\dot{K}_{\rm dim}$ and $\sigma^{2}_{\ddot{\nu}}/\gamma_{\ddot{\nu}}^{2}$. Section~\ref{Sec:Discussion} presents some interesting population-wide trends. Section~\ref{Sec:astro_int} discusses possible astrophysical interpretations of the results of the inference procedure. The conclusions are summarized in Section~\ref{Sec:Conclusions}.

\section{Estimating torque-law parameters} \label{Sec:Methods}

In this section, we review briefly the hierarchical Bayesian scheme proposed by~\citetalias{VargasMelatos2025} to estimate torque-law parameters such as the braking index. The Bayesian scheme ingests pulse times of arrival (TOAs), timing solutions, and their associated uncertainties for a sample of $M$ pulsars. It returns posterior distributions for $n_{\rm pl}+\dot{K}_{\rm dim}$ and $\sigma^{2}_{\ddot{\nu}}/\gamma_{\ddot{\nu}}^{2}$ at both the per-pulsar and population levels. In Section~\ref{subsecII:BrownianModel}, we define the phenomenological Brownian spin-down model introduced by~\citetalias{VargasMelatos2023}~and~\citetalias{VargasMelatos2024}, which combines $\dot{K} \neq 0$ with $\xi(t) \neq 0$ and has been validated with synthetic data through Monte Carlo simulations in the preceding references, e.g.\ for the representative object PSR J0942$-$5552. A key goal of Section~\ref{subsecII:BrownianModel} is to explain the role played by the physical parameters $n_{\rm pl}+\dot{K}_{\rm dim}$ and $\sigma^{2}_{\ddot{\nu}}/\gamma_{\ddot{\nu}}^{2}$ in the context of a pulsar timing experiment. In Section~\ref{subsecII:Bayesframework}, we explain how to combine the Brownian spin-down model with a hierarchical version of Bayes's theorem and a Monte Carlo sampler to calculate the desired posteriors. For the sake of reproducibility, we record explicit formulas for the per-pulsar likelihood, population-level prior, and hyperparameter priors in the hierarchical Bayesian scheme. 

\subsection{Brownian spin-down model} \label{subsecII:BrownianModel}

In the absence of a complete theory describing the internal and radiation reaction torques acting on a rotation-powered pulsar~\citep{AbolmasovBiryukov2024}, we employ an idealized, phenomenological model that combines secular spin down (with $\dot{K}=0$ or $\dot{K}\neq0$) and stochastic spin wandering (i.e. timing noise without glitches). The model assumes that a pulsar spins down secularly due to a torque $\propto K(t)\nu(t)^{n_{\rm pl}}$ [scenario (i) in Section~\ref{Sec:Introduction}], with $n_{\rm pl }\lesssim 7$ for many plausible electromagnetic or gravitational radiation reaction mechanisms (see Section~\ref{Sec:Introduction}). It also executes mean-reverting Brownian motion around the secular spin-down trend due to a stochastic torque [scenario (ii) in Section~\ref{Sec:Introduction}]. Mean reversion leads to a broken power law for the phase residual power spectral density (PSD), as observed~\citep{ParthasarathyShannon2019,LowerBailes2020,GoncharovReardon2021,AntonelliBasu2023}, as well as bounded, long-term values for $\nu, \dot{\nu}$ and $\ddot{\nu}$ satisfying $n \approx n_{\rm pl}+\dot{K}_{\rm dim}$.  For a typical pulsar, the fractional fluctuations in $\nu$, $\dot{\nu}$, and $\ddot{\nu}$ in the Brownian model satisfy $|\delta\nu / \nu | \ll 1$, $| \delta\dot{\nu} / \dot{\nu} | \ll 1$, and $| \delta\ddot{\nu} / \ddot{\nu} | \gtrsim 1$ respectively. The fluctuations in $\ddot{\nu}$ (and hence $n$) time-average to a negligible value for $ T_{\rm obs} \gtrsim~3 \times10^{5} \linebreak \times(\sigma_{\ddot{\nu}}^{2} / 10^{-55} \, {\rm Hz}^2{\rm s}^{-5 }  ) (\gamma_{\ddot{\nu}}/10^{-6} \, {\rm s^{-1}})^{-2}(\dot{\nu} / 10^{-14} \, {\rm Hz \, s^{-1}})^{-4}(\nu / 1 \, {\rm Hz})^{2} \\
\,{\rm years}$ [see equation~(\ref{Eq_SecI:Variance_n})], i.e. observation times much longer than any contemporary pulsar timing campaign. For a comprehensive discussion of the model assumptions, see~\citetalias{VargasMelatos2023} and \citetalias{VargasMelatos2024}. 

In general, the rotational evolution of a pulsar is described by the rotational phase of the crust $\phi(t)$, the frequency $\nu(t) = \dot{\phi}(t)$, and the time derivatives $\dot{\nu}(t)$ and $\ddot{\nu}(t)$. These dynamical variables are packaged in the state vector ${\bf X}=(X_{1},X_{2},X_{3},X_{4})^{\rm T} = (\phi,\nu,\dot{\nu},\ddot{\nu})^{\rm T}$, where ${\rm T}$ denotes the matrix transpose, and evolve according to the stochastic differential equation~\citep{MeyersMelatos2021,MeyersO'Neill2021,AntonelliBasu2023,VargasMelatos2023,O'NeillMeyers2024, VargasMelatos2024, VargasMelatos2025}

\begin{equation}
    d{\bf X}=({\bf A}{\bf X}+{\bf E})dt+{\bf \Sigma} d{\bf B}(t),
    \label{Eq_secII:dX}
\end{equation}

with

\begin{equation}
    \bm A = \begin{pmatrix} 0 & 1 & 0 & 0 \\ 0 & -\gamma_{\nu} & 1 & 0\\ 0 & 0 & -\gamma_{\dot{\nu}} & 1 \\
    0 & 0 & 0 & -\gamma_{\ddot{\nu}} \end{pmatrix}, \label{Eq_secII:Amplitudes_Matrix} 
\end{equation}

\begin{equation}
    \bm E = \begin{pmatrix} 0 \\ \gamma_{\nu} \nu_{\rm pl}(t) \\ \gamma_{\dot{\nu}} \dot{\nu}_{\rm pl}(t) \\ \dddot{\nu}_{\rm pl}(t)+\gamma_{\ddot{\nu}} \ddot{\nu}_{\rm pl}(t) \end{pmatrix},
    \label{Eq_secII:torque_vector_E}
\end{equation}

\noindent and

\begin{equation}
    \bm \Sigma = \text{diag}\left(0, 0, 0 ,\sigma_{\ddot{\nu}} \right). \label{Eq_secII:Sigma_Matrix}
\end{equation}

The $({\bf A}{\bf X}+{\bf E})dt$ term in equation~(\ref{Eq_secII:dX}) is deterministic; it describes mean reversion. In equations~(\ref{Eq_secII:Amplitudes_Matrix})~and~(\ref{Eq_secII:torque_vector_E}), the parameters $\gamma_{\nu}, \gamma_{\dot{\nu}},$ and $\gamma_{\ddot{\nu}}$ are constant damping terms, i.e. the random walk in (say) $\nu(t)$ reverts to the mean $\langle \nu(t) \rangle = \nu_{\rm pl}(t)$ on a time-scale $\gamma_{\nu}^{-1}$ [and similarly for $\dot{\nu}(t)$ and $\ddot{\nu}(t)$]. The mean $\nu_{\rm pl}(t)$ is the solution to the secular braking law $\dot{\nu}_{\rm pl}(t) = K(t)\nu_{\rm pl}(t)^{n_{\rm pl}}$, while $\dot{\nu}_{\rm pl}(t),\ddot{\nu}_{\rm pl}(t),$ and $\dddot{\nu}_{\rm pl}(t)$ correspond to the first, second, and third time derivatives of $\nu_{\rm pl}(t)$, respectively. 

The ${\bf \Sigma} d{\bf B}(t)$ term in equation~(\ref{Eq_secII:dX}) is stochastic. It describes a fluctuating Langevin driver, which represents timing noise. In equation~(\ref{Eq_secII:Sigma_Matrix}), the parameter $\sigma^{2}_{\ddot{\nu}}$ is the squared amplitude of the Langevin driver, which we approximate as a memoryless, white-noise process with

\begin{equation}
    \langle dB_{i}(t)\rangle=0
    \label{Eq_SecII:dB_average}
\end{equation}

and

\begin{equation}
    \langle dB_{i}(t)dB_{j}(t') \rangle=\delta_{ij}\delta(t-t'),
    \label{Eq_SecII:dB_mem_less}
\end{equation} 

where $\langle ...\rangle$ denotes an ensemble average over random realizations of $d{\bf B}(t)$. For simplicity, we assume in equation~(\ref{Eq_secII:Sigma_Matrix}) that the cross-correlation terms vanish, i.e. $\Sigma_{ij}=0$ for $i\neq j$, and set $\Sigma_{11}=\Sigma_{22}=\Sigma_{33}=0$ to ensure ${\bf X}$ is differentiable. The latter property is an important prerequisite to obtain equation~(\ref{Eq_SecI:Variance_n}) theoretically; see Appendix A1 of \citetalias{VargasMelatos2024} for a discussion of this subtle point. However, the model is readily modifiable to include fluctuations in other rotational parameters according to the application at hand, if the data warrant. For example, $\Sigma_{11} \neq 0$ describes noise in $\phi(t)$ due to magnetospheric fluctuations, which perturb the longitude of the pulsar's radio beam. Numerical calculations confirm that astrophysically plausible values for $\gamma_{\nu}, \gamma_{\dot{\nu}}, \gamma_{\ddot{\nu}},$ and $\sigma^{2}_{\ddot{\nu}}$ in the Brownian model, given by equations (\ref{Eq_secII:dX})--(\ref{Eq_SecII:dB_mem_less}), produce synthetic TOAs qualitatively consistent with those observed in real pulsars, e.g. similar root-mean-square and autocorrelation time-scale. The reader is referred to Section~2.4 in~\citetalias{VargasMelatos2024} for a quantitative demonstration of TOA consistency for PSR J0942$-$5552

For a specific random realization of $d{\bf B}(t)$, the analytical solution for ${\bf X}(t)$ (see Appendix A1 in both~\citetalias{VargasMelatos2023} and \citetalias{VargasMelatos2024}) can be used to predict the $n$ value measured between two epochs $t_{1}$ and $t_{2} > t_{1}$ via the standard nonlocal~formula~\citep{JohnstonGalloway1999}\footnote{There is a subtle mathematical distinction between the nonlocal measurement given by equation~(\ref{Eq_SecII:n_meas_nonlocal}) and the local measurement $n(t) = \nu(t)\ddot{\nu}(t)/\dot{\nu}(t)^{2}$ when calculating moments theoretically, as discussed in Appendix A1 of~\citetalias{VargasMelatos2024}.}

\begin{equation}
    n=1-\frac{\dot{\nu}(t_{1})\nu(t_{2})-\dot{\nu}(t_{2})\nu(t_{1})}{\dot{\nu}(t_{1})\dot{\nu}(t_{2})T_{\rm obs}}.
    \label{Eq_SecII:n_meas_nonlocal}
\end{equation}

In the astrophysically relevant regime $\gamma_{\nu} \sim \gamma_{\dot{\nu}} \ll T_{\rm obs}^{-1} \ll \gamma_{\ddot{\nu}}$~\citep{PriceLink2012,MeyersMelatos2021,MeyersO'Neill2021,VargasMelatos2023,O'NeillMeyers2024,VargasMelatos2024}, averaging equation~(\ref{Eq_SecII:n_meas_nonlocal}) over the $d{\bf B}(t)$ ensemble yields $\langle n \rangle = n_{\rm pl}+\dot{K}_{\rm dim}$ and $\langle n^{2} \rangle$ given by equation~(\ref{Eq_SecI:Variance_n}). The formulas for $\langle n\rangle$ and $\langle n^{2} \rangle$ do not depend on the functional form of $K(t)$, as long as the conditions $\vert \dot{K}(t)/K(t)\vert \sim \tau_{K}^{-1}$ and $\tau_{K} \lesssim \tau_{\rm sd}$ hold (see Appendix A2 of \citetalias{VargasMelatos2024}).

\subsection{Hierarchical Bayesian scheme} \label{subsecII:Bayesframework}

The hierarchical Bayesian scheme operates on a population of $M$ pulsars, indexed by $1 \leq m \leq M$. Each pulsar is equipped with a traditional measurement of its braking index $n^{(m)}_{\rm meas}$ [which can be anomalous or not; see equation~(\ref{Eq_SecI:Intro_n_def})], the root-mean-square of its timing residuals $S_{\rm meas}^{(m)}$, and their associated uncertainties $\Delta n^{(m)}_{\rm meas}$ and $\Delta S_{\rm meas}^{(m)}$, respectively. We explain the process to measure $n^{(m)}_{\rm meas},S_{\rm meas}^{(m)},\Delta n^{(m)}_{\rm meas},$ and $\Delta S_{\rm meas}^{(m)}$ from TOA data in Section~\ref{subsecIII:measurements_D^m}. The data for the $m$-th pulsar are denoted by $D^{(m)}=\{n^{(m)}_{\rm meas},S_{\rm meas}^{(m)},\Delta n^{(m)}_{\rm meas},\Delta S_{\rm meas}^{(m)}\}$. The Bayesian scheme ingests the data for all $M$ pulsars, denoted by $D=\{D^{(1)},...,D^{(M)}\}$.

In the Brownian model in Section~\ref{subsecII:BrownianModel}, each pulsar is characterized by the parameters $(n_{\rm pl}+\dot{K}_{\rm dim})^{(m)}$ and $\chi^{(m)}=\sigma^{(m)}_{\ddot{\nu}}[\gamma^{(m)}_{\ddot{\nu}}]^{-1}$. In this paper, we assume that the inference problem is hierarchical. That is, we assume that the pulsars are exchangeable statistically~\citep{gelman2013bayesian,VargasMelatos2025} and that $(n_{\rm pl}+\dot{K}_{\rm dim})^{(m)}$ and $\chi^{(m)}$ are drawn from the population-level prior $\pi[(n_{\rm pl}+\dot{K}_{\rm dim})^{(m)}, \chi^{(m)}\vert \mu_{\rm pl},\sigma_{\rm pl}]$, which is parameterized by the population-level hyperparameters $\mu_{\rm pl}$ and $\sigma_{\rm pl}$~\citepalias{VargasMelatos2025}. Henceforth, we denote the per-pulsar parameters and population-level hyperparameters as $\theta^{(m)}=\{(n_{\rm pl}+\dot{K}_{\rm dim})^{(m)}, \chi^{(m)}\}$ and $\psi = \{\mu_{\rm pl}, \sigma_{\rm pl}\}$, respectively. The set of per-pulsar parameters for all $M$ pulsars is denoted by $\theta = \{\theta^{(1)},...,\theta^{(M)}\}$.

The general form of the $(2M+2)$-dimensional posterior distribution $p(\psi,\theta\vert D)$ is deduced from Bayes's theorem to be

\begin{equation}
    p(\psi, \theta | D) = {\cal Z}^{-1}\prod_{m'=1}^{M} {\cal L}^{(m')}[D^{(m')} \vert \theta^{(m')}]\pi[\theta^{(m')} \vert \psi]\pi(\psi), 
    \label{eq_SecII:BAYES}
\end{equation}

with

\begin{equation}
    {\cal Z}=\int d\psi d\theta\prod_{m'=1}^{M} {\cal L}^{(m')}[D^{(m')} \vert \theta^{(m')}]\pi[\theta^{(m')} \vert \psi]\pi(\psi).
    \label{eq_SecII:Z_BAYES}
\end{equation}

In equations~(\ref{eq_SecII:BAYES}) and (\ref{eq_SecII:Z_BAYES}), ${\cal L}^{(m')}$ denotes the per-pulsar likelihood for the $m'$-th pulsar. It is defined as the product of two factors, one related to measuring $n_{\rm meas}^{(m)}$ and the other related to measuring $S_{\rm meas}^{(m)}$. The first factor is given by 

\begin{equation}
n_{\rm meas}^{(m)} \sim {\cal N}[ (n_{\rm pl}+\dot{K}_{\rm dim})^{(m)} , \{ [\chi^{(m)}s^{(m)}]^2 + [\Delta n_{\rm meas}^{(m)}]^2 \}^{1/2} ],
\label{Eq_secII:nmeas_dist}
\end{equation}

where the notation $X \sim {\cal N}(a,b)$ symbolizes that the random variate $X$ is distributed as a Gaussian with mean $a$ and standard deviation $b$, and $s^{(m)} = \nu^{(m)}[\dot{\nu}^{(m)}]^{-2}[T_{\rm obs}^{(m)}]^{-1/2}$ is a bundle of accurately measured per-pulsar parameters. Equation~(\ref{Eq_secII:nmeas_dist}) is a restatement of the rightmost term of equation~(\ref{Eq_SecI:Variance_n}) with measurement errors added. The second factor is approximately given by

\begin{equation}
\log S_{\rm meas}^{(m)} \sim {\cal N}[\mu_{ S \rm, BM}[\chi^{(m)}] , \{\sigma^{2}_{S \rm, BM} +[\Delta S_{\rm meas}^{(m)}]^{2}\}^{1/2} ],
\label{Eq_secII:Smeas_dist}
\end{equation}

where the log-normal is in base $e$. Equation~(\ref{Eq_secII:Smeas_dist}) is an approximation. It follows from analytical calculations and simulations involving synthetic data as demonstrated in Appendix B~of~\citetalias{VargasMelatos2025}. The arguments of ${\cal N}(a,b)$ in equation~(\ref{Eq_secII:Smeas_dist}) and their associated calibration are discussed in Appendix~\ref{App:S_m_calibration}.

In this paper, noting the paucity of first-principles knowledge about the astrophysical distribution of $(n_{\rm pl}+\dot{K}_{\rm dim})^{(m)}$ and $\chi^{(m)}$~\citep{AbolmasovBiryukov2024}, we set the population-level prior $\pi[\theta^{(m)}\vert \psi]$ to

\begin{equation}
    (n_{\rm pl}+\dot{K}_{\rm dim})^{(m)} \sim {\cal N}(\mu_{\rm pl}, \sigma_{\rm pl}).
    \label{eq_SecII:npl_Kdim_dist}
\end{equation}

In writing down equation~(\ref{eq_SecII:npl_Kdim_dist}), we assume that the $M$ pulsars analyzed here spin down secularly via the same mechanism, e.g. electromagnetic or gravitational radiation reaction (see Section~\ref{Sec:Introduction}). This astrophysical scenario may not hold in reality of course  --- indeed, alternatives are discussed in Section~\ref{Sec:astro_int} --- and equation (\ref{eq_SecII:npl_Kdim_dist}) can be changed if the need arises. Additionally, the population-level prior~(\ref{eq_SecII:npl_Kdim_dist}) is independent of $\chi^{(m)}$, as there is no compelling astrophysical reason to prefer particular values of $\chi^{(m)}$ over others. So far, $\gamma_{\ddot{\nu}}$ has only been measured in a few pulsars through timing noise auto-correlation time-scales~\citep{PriceLink2012,O'NeillMeyers2024}, while $\sigma_{\ddot{\nu}}$ is known to span multiple decades across the pulsars whose timing noise amplitudes have been measured~\citep{CordesHelfand1980,ColesHobbs2011,ParthasarathyShannon2019,LowerBailes2020,GoncharovReardon2021}. To accommodate the latter property, we choose the loose prior 

\begin{equation}
    \log \chi^{(m)} \sim {\cal N}(-19.5,5),
    \label{eq_SecII:prior_chi}
\end{equation}

to cover many decades of plausible $\chi^{(m)}$ values approximately uniformly, while ensuring numerical convergence. The prior on $\chi^{(m)}$ is centered at the typical values of $\sigma_{\ddot{\nu}}$ and $\gamma_{\ddot{\nu}}$, where anomalous braking indices $\vert n \vert \gg n_{\rm pl}$ are obtained~\citepalias{VargasMelatos2023,VargasMelatos2024}.

The prior distribution of the hyperparameters $\pi(\psi)$ is set to be consistent with the astrophysical scenario in the previous paragraph. We assume that $\mu_{\rm pl}$ and $\sigma_{\rm pl}$ are statistically independent, i.e. $\pi(\psi)=\pi(\mu_{\rm pl})\pi(\sigma_{\rm pl})$. We also set

\begin{equation}
    \mu_{\rm pl} \sim {\cal U}(-75,75),
    \label{eq_SecII:prior_mupl}
\end{equation}

where $X \sim {\cal U}(a,b)$ denotes that $X$ is uniformly distributed on the domain $a \leq X \leq b$. Finally, we set

\begin{equation}
    \log \sigma_{\rm pl} \sim {\cal N}(0,2).
    \label{eq_secII:prior_sigmapl}
\end{equation}

When analyzing real data in Section~\ref{Sec:Analysis}, we verify a posteriori that the results are insensitive to a wider domain choice than in equations~(\ref{eq_SecII:prior_chi})--(\ref{eq_secII:prior_sigmapl}), e.g. $\log \chi^{(m)} \sim {\cal N}(-5,20)$ or $-10^{3} \leq \mu_{\rm pl} \leq 10^{3}$. Note that equation (\ref{eq_SecII:prior_mupl}) is conservative; it extends well beyond the traditional range $2 \lesssim \mu_{\rm pl} \leq 7$ associated with electromagnetic and gravitational radiation reaction to encompass other mechanisms, e.g.\ Hall waves generated in the pulsar's superfluid core and/or crust~\citep{BransgroveLevin2025}.

The hierarchical Bayesian scheme defined by equations (\ref{eq_SecII:BAYES})--(\ref{eq_secII:prior_sigmapl}) is implemented in \texttt{Stan}.\footnote{\href{https://github.com/stan-dev}{https://github.com/stan-dev}} We use a Hamiltonian Monte Carlo No U-Turn Sampler \citep{Betancourt2017}~to compute the full posterior $p(\psi,\theta\vert D)$ [equation~(\ref{eq_SecII:BAYES})]. To assist with numerical convergence, we truncate equation~(\ref{eq_SecII:npl_Kdim_dist}) to the same domain as the $\mu_{\rm pl}$ prior. Having computed $p(\psi,\theta \vert D)$, we compute the posterior distribution for the hyperparameters by marginalizing over $\theta^{(1)}, \dots, \theta^{(M)}$, viz.

\begin{equation}
    p(\psi \vert D) = \int d\theta~p(\psi,\theta \vert D).
    \label{eq_secII:p_hyperparams}
\end{equation}

We also compute the population-level posterior distribution of $n_{\rm pl}+\dot{K}_{\rm dim}$ (which amounts to a posterior predictive check) according to

\begin{align} 
p[(n_{\rm pl}+\dot{K}_{\rm dim})^{({\rm pop})} | D] =& \int d\mu_{\rm pl} \, d\sigma_{\rm pl} \, \nonumber \\ &\times\pi[ (n_{\rm pl}+\dot{K}_{\rm dim})^{({\rm pop})} | \mu_{\rm pl}, \sigma_{\rm pl} ] \nonumber \\
&\times p(\mu_{\rm pl},\sigma_{\rm pl} | D). 
\label{Eq_subsecII:ppc_n_pop}
\end{align}

In the second line of equation~(\ref{Eq_subsecII:ppc_n_pop}), we substitute the symbol $(n_{\rm pl}+\dot{K}_{\rm dim})^{({\rm pop})}$ to replace $(n_{\rm pl}+\dot{K}_{\rm dim})^{(m)}$ in the analytical form of the population-level prior, i.e. $(n_{\rm pl}+\dot{K}_{\rm dim})^{(m)} \mapsto (n_{\rm pl}+\dot{K}_{\rm dim})^{({\rm pop})}$ in equation~(\ref{eq_SecII:npl_Kdim_dist}). To assist the reader, Table~\ref{Table_subsecII:priorsHB} summarizes the priors used throughout the paper for the population-level hyperparameters $\psi$ and the per-pulsar parameters $\theta^{(m)}$.

\begin{table}
\centering
\caption{Prior ranges used by the hierarchical Bayesian analysis in Section~\ref{Sec:Analysis} for the population-level hyperparameters $\psi$ (upper half) and the per-pulsar parameters $\theta^{(m)}$ (lower half; identical for $1\leq m \leq M$). The last two columns parametrize ${\cal U}(a,b)$ or ${\cal N}(a,b)$ according to the prior used (third column).}
\label{Table_subsecII:priorsHB}
\begin{tabular*}{\columnwidth}{p{0.25\columnwidth}p{0.1\columnwidth}p{0.2\columnwidth}p{0.15\columnwidth}p{0.15\columnwidth}}
\hline
Parameter & Units & Prior & $a$ & $b$ \\
\hline
 $\mu_{\rm pl}$ & --- & Uniform & $-75$ & $75$ \\
 $\sigma_{\rm pl}$ & --- & Log-normal & $0$ & $2$\\
 \hline 
 $(n_{\rm pl}+\dot{K}_{\rm dim})^{(m)}$ & --- & Normal & $\mu_{\rm pl}$ & $\sigma_{\rm pl}$ \\
 $\chi^{(m)}$ & s$^{-5/2}$ & Log-normal & $-19.5$ & $5$ \\
\hline
\end{tabular*}
\end{table}

\section{Data} \label{Sec:Observations}

In this section, we discuss briefly the sample of pulsars analyzed in this paper, and the recipe to generate the data, $D$, ingested by the Bayesian scheme. Section~\ref{subsecIII:pulsar_sample} provides an overview of the 85 objects analyzed originally by~\cite{ParthasarathyShannon2019} and justifies restricting the analysis in this paper to a subsample of $68$ objects. In Section~\ref{subsecIII:measurements_D^m}, we discuss how to convert TOA data into per-pulsar measurements $D^{(m)}=\{n_{\rm meas}^{(m)}, \Delta n_{\rm meas}^{(m)},S_{\rm meas}^{(m)},\Delta S_{\rm meas}^{(m)}\}$, which are fed into the hierarchical Bayesian scheme set out in Section~\ref{subsecII:Bayesframework}. 

\subsection{Pulsar sample} \label{subsecIII:pulsar_sample}

In this paper, we analyze a subsample of pulsars studied initially by~\cite{ParthasarathyShannon2019} and~\cite{ParthasarathyJohnston2020}. These pulsars were observed monthly using Murriyang, the 64-m CSIRO Parkes Radio Telescope, in support of the Fermi Gamma-Ray Space Telescope mission~\citep{SmithGuillemot2008} and as part of the young pulsar timing program (P574)~\citep{WeltevredeJohnston2010,JohnstonSobey2021}. 

The sample was observed using the 20-cm multibeam receiver~\citep{Staveley-SmithWilson1996}, which operates over a bandwidth of $256$ MHz divided into 1024 frequency channels. The measured voltage time-series was folded in real time to form {\sc PSRFITS}~\citep{HotanvanStraten2004} format folded archives with 1024 phase bins. These archives were then processed, calibrated and cleaned following the steps outlined in~\cite{ParthasarathyShannon2019}, before being averaged in frequency, time and polarization to form one-dimensional total intensity profiles. The TOAs of the pulses are obtained by cross-correlating a smoothed, noise-free template profile against the total intensity pulse profiles.

In this paper, we analyze a subsample of $68$ pulsars selected from the original 85 pulsars~in~\cite{ParthasarathyShannon2019}. The $17$ excluded pulsars are divided into two categories. The first category contains four objects, PSR J1632$-$4818, PSR J1638$-$4608,  PSR J1715$-$3903, and PSR J1830$-$1059, which glitched during the P574 observation campaign.\footnote{Glitch parameters are summarized in \\
\href{https://www.jb.man.ac.uk/pulsar/glitches/gTable.html}{https://www.jb.man.ac.uk/pulsar/glitches/gTable.html}.} The second category contains $13$ objects which exhibit timing residuals qualitatively consistent with noise in the rotational phase $(\sigma_{\phi} = \Sigma_{11}  \neq 0)$. Glitches and phase noise are excluded from the simplest version of the Brownian model in Section~\ref{subsecII:BrownianModel}. For completeness, however, we make a first attempt at generalizing the analysis to include the $13$ objects consistent with $\sigma_{\phi} \neq 0$ in Appendix~\ref{App:analysis_all_PSRs}.

\subsection{Per-pulsar data $D^{(m)}$}\label{subsecIII:measurements_D^m}

The raw data $D=\{n_{\rm meas}^{(m)}, \Delta n_{\rm meas}^{(m)},S_{\rm meas}^{(m)},\Delta S_{\rm meas}^{(m)}\}_{1\leq m \leq M=81}$ for the 81 non-glitching pulsars in the full~\cite{ParthasarathyShannon2019} sample are tabulated in Appendix C for the sake of reproducibility; see Table~\ref{AppB:TableD^m}.

The recipe to generate the data $D^{(m)}$ for the $m$-th pulsar proceeds as follows. We fit a standard second-order polynomial ephemeris [up to $\ddot{\nu}^{(m)}$] to the $N_{\rm TOA}^{(m)}$ TOAs, $\{t_{1}^{(m)},\dots, t^{(m)}_{N^{(m)}_{\rm TOA}}\}$, spanning the observation time $T_{\rm obs}^{(m)}$. In practice, we use~\tempoDOS~\citep{HobbsEdwards2006}~to fit the TOAs, but other timing software can be used if preferred by calibrating the arguments of equation~(\ref{Eq_secII:Smeas_dist}) as discussed in Appendix~\ref{App:S_m_calibration}. We employ~\tempoDOS~because the condition $\Delta n_{\rm meas} \propto \Delta \ddot{\nu} \lesssim \chi s$ holds regardless and other traditional methods, such as~\temponest~\citep{LentatiAlexander2014}, incur significantly higher computational costs (see Appendix~\ref{App:uncF2}). From the polynomial fit, the values $\nu^{(m)}, \dot{\nu}^{(m)},\ddot{\nu}^{(m)},$ and $\Delta\ddot{\nu}^{(m)}$, the timing residuals $\{{\cal R}[t_{1}^{(m)}],\dots,{\cal R}[t^{(m)}_{N^{(m)}_{\rm TOA}}]\}$, and their associated uncertainties $\{\Delta{\cal R}[t_{1}^{(m)}],\dots,\Delta{\cal R}[t^{(m)}_{N^{(m)}_{\rm TOA}}]\}$ are obtained. We then compute $n_{\rm meas}^{(m)}=\nu^{(m)}\ddot{\nu}^{(m)}[\dot{\nu}^{(m)}]^{-2}$ and $\Delta n_{\rm meas}^{(m)} = \nu^{(m)}\Delta \ddot{\nu}^{(m)}[\dot{\nu}^{(m)}]^{-2}$, directly.\footnote{The uncertainties $\Delta \nu$ and $\Delta \dot{\nu}$ satisfy $\Delta \nu/\nu\sim 10^{-9}$ and $\Delta \dot{\nu}/\dot{\nu}\sim10^{-3}$, so $\Delta n_{\rm meas}$ is dominated by $\Delta \ddot{\nu}$; see Section~2.3 in \citetalias{VargasMelatos2023}.} 

The root-mean-square $S_{\rm meas}^{(m)}$ of the timing residuals is computed from

\begin{equation}
    S^{(m)}_{\rm meas} = \sqrt{\frac{1}{N_{\rm TOA}^{(m)}}\sum_{i=1}^{N_{\rm TOA}^{(m)}} {\cal R}[t_{i}^{(m)}]^{2} }.\label{eq_subsecIII:S_meas_m}
\end{equation}

The associated uncertainty $\Delta S_{\rm meas}^{(m)}$ is estimated by a bootstrap method, developed by \citetalias{VargasMelatos2025}, which assumes that the $i$-th timing residual, with $1 \leq i \leq N_{\rm TOA}^{(m)}$, is drawn from the distribution ${\cal N}\{{\cal R}[t_{i}^{(m)}], \Delta{\cal R}[t_{i}^{(m)}] \}$. The latter distribution is used to generate $10^{5}$ new ${\cal R}[t_{i}^{(m)}]$ samples, for each $i$-value, which in turn are combined into $10^{5}$ new $S_{\rm meas}^{(m)}$ values through equation~(\ref{eq_subsecIII:S_meas_m}). The standard deviation of the $10^{5}$ new $S_{\rm meas}^{(m)}$ values is $\Delta S_{\rm meas}^{(m)}$.      

\section{Posterior distributions} \label{Sec:Analysis}

In this section we present the inference results from applying the Brownian model in Section~\ref{subsecII:BrownianModel} and the hierarchical Bayesian scheme in Section~\ref{subsecII:Bayesframework} to the sample of $68$ pulsars specified in Section~\ref{subsecIII:pulsar_sample} and Appendix~\ref{subsec_AppB:D^m_for_68_psrs}. Specifically, we present the joint posterior distribution for the population-level hyperparameters $\mu_{\rm pl}$ and $\sigma_{\rm pl}$ in Section~\ref{subsecIV:hp_results} and for the per-pulsar Brownian model parameters $\theta^{(m)}$ in Section~\ref{subsecIV:pp_results}.

\subsection{Population-level hyperparameters} \label{subsecIV:hp_results}

\begin{figure}
\flushleft
 \includegraphics[width=\columnwidth]{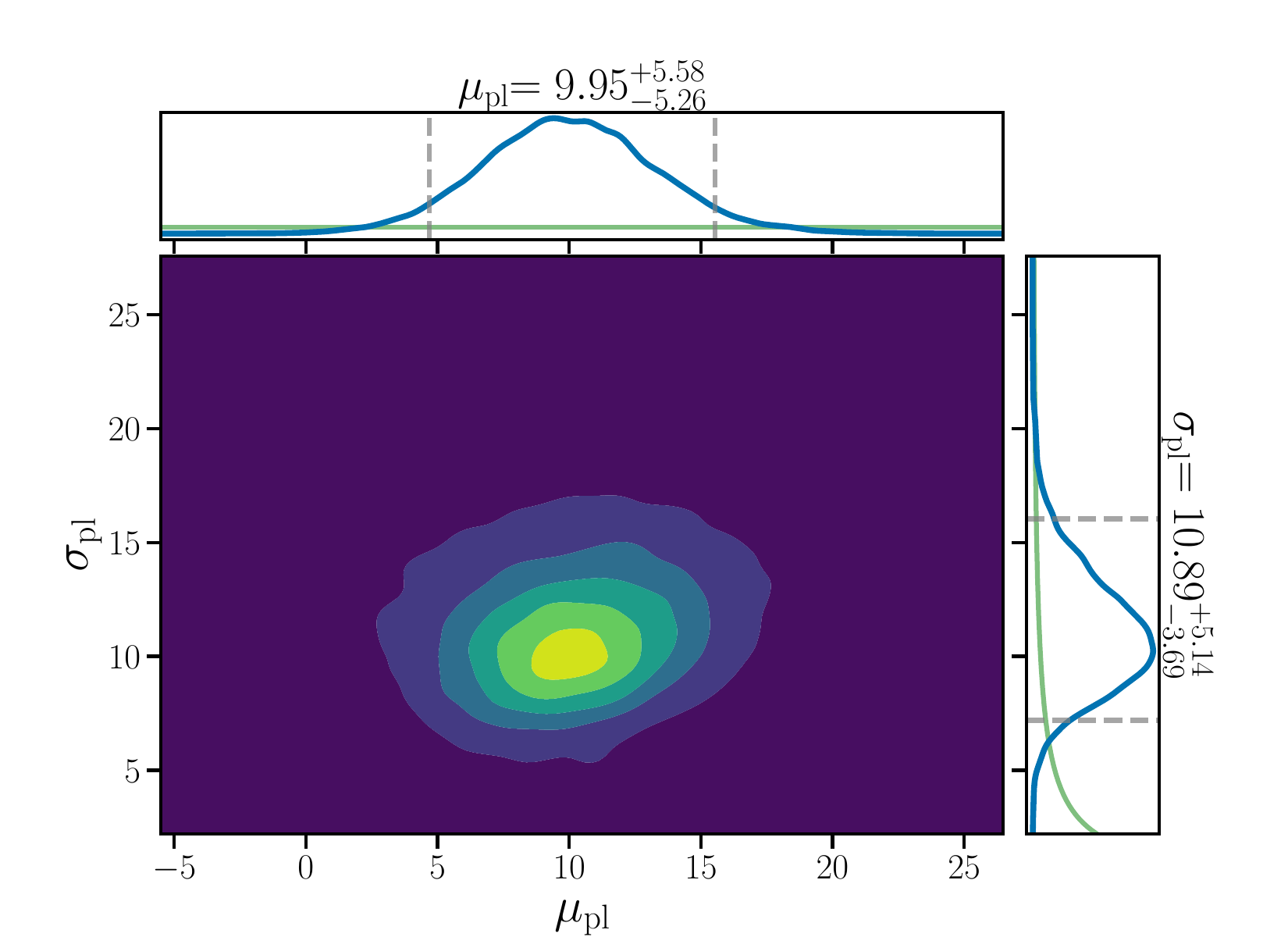}
 \caption{Posterior distribution $p(\mu_{\rm pl}, \sigma_{\rm pl} \vert D)$ (central panel) for the population-level hyperparameters $\mu_{\rm pl}$ and $\sigma_{\rm pl}$ for a subsample of P574 pulsars ($M=68$; see Table~\ref{AppB:TableD^m}). The contours of $p(\mu_{\rm pl}, \sigma_{\rm pl} \vert D)$ mark the $30\%,50\%$,$70\%$, and $90\%$ credible intervals. The one-dimensional posteriors correspond to $p(\mu_{\rm pl}, \sigma_{\rm pl} \vert D)$ marginalized over $\mu_{\rm pl}$ (right panel) and $\sigma_{\rm pl}$ (upper panel). For both the upper and right panels, the marginalized posteriors are plotted as blue curves, while the priors are plotted as green curves. The surtitles for the upper and right panels display the corresponding inferred median (central value) and $90\%$ credible interval (error bars; grey dashed lines).}
\label{fig_secIV:corner_plot}
\end{figure}

Fig.~\ref{fig_secIV:corner_plot} displays the posterior distribution $p(\mu_{\rm pl}, \sigma_{\rm pl} \vert D)$ [equation~(\ref{eq_secII:p_hyperparams})] as a traditional corner plot. The central panel graphs the function $p(\mu_{\rm pl},\sigma_{\rm pl}|D)$ as a contour plot on the $\mu_{\rm pl}$-$\sigma_{\rm pl}$ plane, where the contours mark the $30\%, 50\%, 70\%$, and $90\%$ credible intervals. The upper and right panels show the one-dimensional marginalized posteriors for $\mu_{\rm pl}$ and $\sigma_{\rm pl}$, respectively. The central values for $\mu_{\rm pl}$ and $\sigma_{\rm pl}$ --- quoted on top of their respective panels --- correspond to the posterior medians, while the error bars define the $90\%$ credible interval. The error bars are bracketed by dashed lines in the one-dimensional posteriors. The green lines in the upper and right panels represents the priors used for $\mu_{\rm pl}$ and $\sigma_{\rm pl}$, respectively, within the plotted domain (see Table~\ref{Table_subsecII:priorsHB}).

The posterior in Fig.~\ref{fig_secIV:corner_plot} is unimodal. There is no evidence of railing against the (generous) prior bounds, and the sampler does not return any warnings about numerical convergence, consistent with Monte Carlo validation tests performed for $M \geq 50$ by~\citetalias{VargasMelatos2025}. The parameter estimates $\mu_{\rm pl} = 9.95^{+5.58}_{-5.26}$ and $\sigma_{\rm pl} = 10.89^{+5.14}_{-3.69}$ are broadly reasonable, given what is known about the relative magnitudes of the contributions of timing noise and secular evolution in anomalous braking indices in other objects, e.g.\ PSR J0942$-$5552~\citepalias{VargasMelatos2023,VargasMelatos2024}. Interestingly, the error bars on $\mu_{\rm pl}$ exclude $n_{\rm pl} = 3$ and $\dot{K}=0$ uniformly across the sample at 90\% confidence. That is, there is statistical evidence that some mechanism other than constant-$K$ magnetic dipole braking (e.g.\ $\dot{K} \neq 0$) operates in a significant fraction of the objects in the sample, after one corrects hierarchically for the (often dominant) stochastic anomaly in $n_{\rm meas}^{(m)}$ caused by $\sigma_{\ddot{\nu}} \neq 0$. The reasonableness and statistical and physical interpretations of the results in Fig.~\ref{fig_secIV:corner_plot} are discussed quantitatively in Sections~\ref{Sec:Discussion} and~\ref{Sec:astro_int}.

\begin{figure}
\flushleft
 \includegraphics[width=\columnwidth]{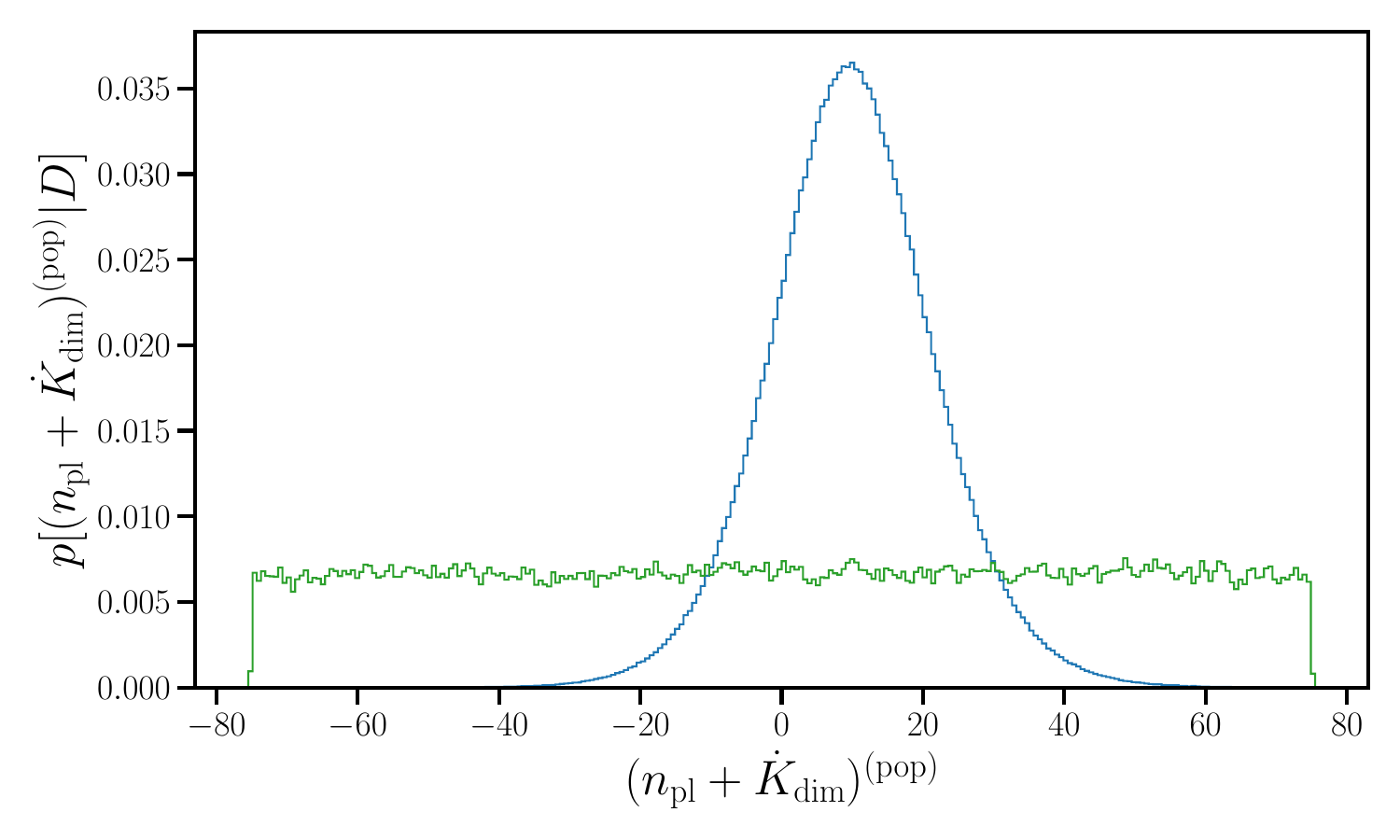}
 \caption{Posterior predictive check: distribution of $n_{\rm pl}+\dot{K}_{\rm dim}$ values without distinguishing between the $68$ exchangeable objects in the analysis, given by equation~(\ref{Eq_subsecII:ppc_n_pop}). The posterior distribution $p[(n_{\rm pl}+\dot{K}_{\rm dim})^{({\rm pop})} \vert D]$ (blue histogram) is obtained by combining the inferred $p(\mu_{\rm pl},\sigma_{\rm pl}\vert D)$ and the population-level prior $\pi[(n_{\rm pl}+\dot{K}_{\rm dim})^{({\rm pop})} \vert \mu_{\rm pl}, \sigma_{\rm pl}]$ via equation~(\ref{Eq_subsecII:ppc_n_pop}). The population-level prior $\pi[(n_{\rm pl}+\dot{K}_{\rm dim})^{(\rm pop)} \vert \psi ]$ [equation~(\ref{eq_SecII:npl_Kdim_dist})] is plotted as a green histogram. }
\label{fig_secIV:ppc}
\end{figure}

Fig.~\ref{fig_secIV:ppc} visualizes the results in Fig.~\ref{fig_secIV:corner_plot} in another, complementary way, through a posterior predictive check. The posterior distribution $p[(n_{\rm pl}+\dot{K}_{\rm dim})^{(\rm pop)}\vert D]$ (blue histogram) is constructed from $p(\mu_{\rm pl}, \sigma_{\rm pl}\vert D)$ via equation~(\ref{Eq_subsecII:ppc_n_pop}). The function $p[(n_{\rm pl}+\dot{K}_{\rm dim})^{(\rm pop)}\vert D]$ represents the distribution of $n_{\rm pl}+\dot{K}_{\rm dim}$ values predicted by the inference scheme across the population, without distinguishing between the $68$ objects in the sample, which are exchangeable statistically~\citep{gelman2013bayesian}. For reference, the green histogram represents the population-level prior $\pi[(n_{\rm pl}+\dot{K}_{\rm dim})^{(\rm pop)} \vert \psi ]$ [equation~(\ref{eq_SecII:npl_Kdim_dist})]. 

\subsection{Per-pulsar Brownian model parameters} \label{subsecIV:pp_results}

The hierarchical Bayesian scheme in Section~\ref{subsecII:Bayesframework} generates a posterior distribution for the Brownian model parameters $\theta^{(m)}$ of the $m$-th pulsar by marginalizing $p(\psi,\theta|D)$ in equation~(\ref{eq_SecII:BAYES}) over $\theta^{(m')}$ for all $m' \neq m$. The properties of the resulting one-dimensional posteriors for $\theta^{(m)}= \{(n_{\rm pl}+\dot{K}_{\rm dim})^{(m)},\chi^{(m)}\}$ ($1\leq m \leq 68$) are quoted in Table~\ref{Tab:per-pulsar_inf}. The reported central values and error bars correspond to the median and $90\%$ credible intervals of the respective one-dimensional posteriors $p[(n_{\rm pl}+\dot{K}_{\rm dim})^{(m)} \vert D]$ (second column) and $p[\chi^{(m)} \vert D]$ (third column). The inferred central values span the ranges $-13.86 \leq (n_{\rm pl}+\dot{K}_{\rm dim})^{(m)} \leq 30.38$, and $-21.20 \leq \log_{10}\chi^{(m)}/(1~{\rm s}^{5/2}) \leq -18.26$. Objects in the top half of Table~\ref{Tab:per-pulsar_inf} have braking indices measured by~\cite{ParthasarathyJohnston2020}, recorded in the fifth column

Table~\ref{Tab:per-pulsar_inf} also presents the standard deviation $\chi^{(m)}s^{(m)}$ (fourth column) for the $n_{\rm meas}^{(m)}$-likelihood in equation~(\ref{Eq_secII:nmeas_dist}). The reader is invited to compare the numbers in the second and fourth columns. Values of $\chi^{(m)}s^{(m)} \gg 1$ imply $\langle [n^{(m)}]^{2} \rangle \gg 1$, i.e. the measured $n_{\rm meas}^{(m)}$ is likely to be anomalous, realizing scenario (ii) in Section~\ref{Sec:Introduction}. They also imply $\chi^{(m)} s^{(m)} \gg n_{\rm pl}^{(m)} + \dot{K}_{\rm dim}^{(m)}$ for many plausible secular braking mechanisms (see Section~\ref{Sec:Introduction}), i.e. the braking index anomaly is predominantly stochastic, as modeled by the mean-reverting random walk in Section~\ref{subsecII:BrownianModel}, rather than secular. The entries for $\chi^{(m)}s^{(m)}$ are calculated by combining the central value of $\chi^{(m)}$ (third column) with the per-pulsar parameter bundle $s^{(m)}$. As $s^{(m)}$ is known to $\lesssim 1\%$ accuracy, the uncertainties on $\chi^{(m)}s^{(m)}$ and $\chi^{(m)}$ are approximately equal. The reasonableness and physical interpretation of the per-pulsar results in Table~\ref{Tab:per-pulsar_inf} are discussed in Sections~\ref{Sec:Discussion} and~\ref{Sec:astro_int}.

\begin{table*}
\begin{center}
\caption{Inferred per-pulsar parameters $\theta^{(m)} = \{(n_{\rm pl}+\dot{K}_{\rm dim})^{(m)},\chi^{(m)}\}$ and $\chi^{(m)}s^{(m)}$ [see equation~(\ref{Eq_secII:nmeas_dist})] for the 68 P574 pulsars analyzed in this paper. The second and third columns display the inferred median (central value) and $90\%$ credible intervals (error bars) of the corresponding one-dimensional posterior. The fourth column displays the median of $\chi^{(m)}s^{(m)}$, noting that $s^{(m)}=[\nu^{(m)}] [\dot{\nu}^{(m)}]^{-2} [T_{\rm obs}^{(m)}]^{-1/2}$ is known with a precision of $1\%$ or better. The pulsars listed in the upper half of the table (delimited by a horizontal line) have a braking index measurement reported by~\protect \cite{ParthasarathyJohnston2020} (last column).}
\label{Tab:per-pulsar_inf}
\renewcommand{\arraystretch}{1.65}  
\begin{tabular*}{\textwidth}{@{\extracolsep{\fill}}lcccc}
\hline
PSR & $n_{\rm pl}+\dot{K}_{\rm dim}$ & $\log_{10} \chi /(1~{\rm s}^{5/2})$ & $ \chi s$ & $n$ \\ 
\hline
J0857$-$4424 & $10.3^{+19.6}_{-19.3}$ & $-18.6^{+0.2}_{-0.2}$ & $1.18\times10^{3}$ & $2890.0^{+30.0}_{-30.0}$  \\
  J0954$-$5430 & $14.5^{+4.0}_{-4.7}$ & $-21.0^{+0.4}_{-0.4}$ & $2.24\times10^{0}$ & $18.0^{+9.0}_{-9.0}$  \\
 J1412$-$6145 & $12.2^{+8.2}_{-9.0}$ & $-19.5^{+0.4}_{-0.4}$ & $5.27\times10^{0}$ & $20.0^{+3.0}_{-3.0}$  \\
 J1509$-$5850 & $11.2^{+2.5}_{-2.6}$ & $-20.5^{+0.4}_{-0.4}$ & $1.22\times10^{0}$ & $11.0^{+3.0}_{-3.0}$  \\
 J1513$-$5908 & $2.83^{+0.01}_{-0.01}$ & $-19.2^{+0.4}_{-0.3}$ & $3.31\times10^{-3}$ & $2.82^{+0.06}_{-0.06}$  \\
 J1524$-$5706 & $4.19^{+0.59}_{-0.56}$ & $-21.2^{+0.4}_{-0.4}$ & $2.88\times10^{-1}$ & $4.2^{+0.7}_{-0.7}$  \\
 J1531$-$5610 & $17.4^{+20.8}_{-18.5}$ & $-18.9^{+0.3}_{-0.3}$ & $2.27\times10^{1}$ & $43.0^{+1.0}_{-1.0}$  \\
 J1637$-$4642 & $30.4^{+8.4}_{-17.8}$ & $-19.0^{+0.4}_{-0.4}$ & $5.71\times10^{0}$ & $34.0^{+3.0}_{-3.0}$  \\
 J1643$-$4505 & $12.6^{+13.1}_{-13.7}$ & $-19.8^{+0.3}_{-0.4}$ & $1.14\times10^{1}$ & $15.0^{+6.0}_{-6.0}$  \\
 J1648$-$4611 & $14.0^{+17.4}_{-17.1}$ & $-19.3^{+0.3}_{-0.3}$ & $2.17\times10^{1}$ & $40.0^{+10.0}_{-10.0}$  \\
 J1738$-$2955 & $8.35^{+19.10}_{-19.40}$ & $-19.0^{+0.3}_{-0.3}$ & $8.38\times10^{1}$ & $-70.0^{+40.0}_{-40.0}$  \\
 J1806$-$2125 & $6.76^{+18.20}_{-16.40}$ & $-19.3^{+0.4}_{-0.3}$ & $2.32\times10^{1}$ & $90.0^{+60.0}_{-60.0}$  \\
 J1809$-$1917 & $20.4^{+9.3}_{-13.5}$ & $-18.9^{+0.4}_{-0.4}$ & $6.77\times10^{0}$ & $23.5^{+6.0}_{-6.0}$  \\
 J1815$-$1738 & $9.6^{+5.6}_{-5.3}$ & $-19.5^{+0.4}_{-0.4}$ & $3.00\times10^{0}$ & $9.0^{+3.0}_{-3.0}$  \\
 J1824$-$1945 & $10.1^{+20.2}_{-18.7}$ & $-19.4^{+0.4}_{-0.4}$ & $3.92\times10^{2}$ & $120.0^{+20.0}_{-20.0}$  \\
 J1833$-$0827 & $-13.9^{+6.1}_{-2.7}$ & $-20.3^{+0.4}_{-0.4}$ & $1.74\times10^{0}$ & $-15.0^{+2.0}_{-2.0}$  \\
 \hline
 J0543$+$2329 & $5.17^{+14.50}_{-12.50}$ & $-20.5^{+0.3}_{-0.3}$ & $1.12\times10^{1}$ & $-$  \\
 J0745$-$5353 & $10.8^{+19.7}_{-18.8}$ & $-20.7^{+0.3}_{-0.3}$ & $2.02\times10^{2}$ & $-$  \\
 J0820$-$3826 & $9.8^{+19.7}_{-18.9}$ & $-19.8^{+0.4}_{-0.4}$ & $2.84\times10^{2}$ & $-$  \\
 J0834$-$4159 & $5.66^{+18.50}_{-17.30}$ & $-20.4^{+0.3}_{-0.3}$ & $2.35\times10^{1}$ & $-$  \\
 J0905$-$5127 & $9.33^{+19.80}_{-19.30}$ & $-19.3^{+0.4}_{-0.3}$ & $1.86\times10^{2}$ & $-$  \\
 J1043$-$6116 & $11.7^{+20.3}_{-18.5}$ & $-20.3^{+0.3}_{-0.3}$ & $6.12\times10^{1}$ & $-$  \\
 J1115$-$6052 & $9.6^{+19.8}_{-19.2}$ & $-20.2^{+0.4}_{-0.4}$ & $1.23\times10^{2}$ & $-$  \\
 J1123$-$6259 & $10.0^{+19.8}_{-19.4}$ & $-19.4^{+0.4}_{-0.3}$ & $1.68\times10^{3}$ & $-$  \\
 J1156$-$5707 & $9.55^{+19.80}_{-19.30}$ & $-18.7^{+0.3}_{-0.3}$ & $3.97\times10^{2}$ & $-$  \\
 J1224$-$6407 & $9.26^{+19.20}_{-18.70}$ & $-20.6^{+0.4}_{-0.4}$ & $5.78\times10^{1}$ & $-$  \\
 J1305$-$6203 & $5.06^{+14.60}_{-12.70}$ & $-20.5^{+0.3}_{-0.4}$ & $1.20\times10^{1}$ & $-$  \\
 J1349$-$6130 & $9.62^{+20.10}_{-18.90}$ & $-19.7^{+0.4}_{-0.3}$ & $6.61\times10^{2}$ & $-$  \\
 J1512$-$5759 & $8.7^{+19.3}_{-18.9}$ & $-19.6^{+0.4}_{-0.3}$ & $6.83\times10^{1}$ & $-$  \\
 J1514$-$5925 & $10.0^{+20.0}_{-18.7}$ & $-19.6^{+0.4}_{-0.4}$ & $6.74\times10^{2}$ & $-$  \\
 J1515$-$5720 & $10.1^{+19.8}_{-18.9}$ & $-20.1^{+0.4}_{-0.3}$ & $2.60\times10^{2}$ & $-$  \\
 J1530$-$5327 & $8.68^{+19.50}_{-19.20}$ & $-20.7^{+0.3}_{-0.3}$ & $1.11\times10^{2}$ & $-$  \\
 J1539$-$5626 & $10.2^{+19.6}_{-19.0}$ & $-19.5^{+0.4}_{-0.3}$ & $1.12\times10^{3}$ & $-$  \\
 J1543$-$5459 & $10.5^{+20.2}_{-18.8}$ & $-19.2^{+0.4}_{-0.3}$ & $7.57\times10^{1}$ & $-$  \\
 J1548$-$5607 & $11.0^{+19.8}_{-18.4}$ & $-19.6^{+0.4}_{-0.3}$ & $5.37\times10^{1}$ & $-$  \\
 J1549$-$4848 & $11.5^{+20.2}_{-18.9}$ & $-19.8^{+0.3}_{-0.3}$ & $1.06\times10^{2}$ & $-$  \\
 J1551$-$5310 & $11.4^{+20.9}_{-19.3}$ & $-18.3^{+0.3}_{-0.3}$ & $7.49\times10^{1}$ & $-$  \\
\hline
\end{tabular*}
\end{center}
\end{table*}

\begin{table*}
\begin{center}
\contcaption{}
\renewcommand{\arraystretch}{1.5}  
\begin{tabular*}{\textwidth}{@{\extracolsep{\fill}}lcccc}
\hline
PSR & $n_{\rm pl}+\dot{K}_{\rm dim}$ & $\log \chi /(1~{\rm s}^{5/2})$ & $\chi s$ & $n$ \\ 
\hline
 J1600$-$5751 & $9.97^{+20.20}_{-19.00}$ & $-19.9^{+0.4}_{-0.3}$ & $1.06\times10^{3}$ & $-$  \\
 J1601$-$5335 & $10.8^{+16.2}_{-15.9}$ & $-19.3^{+0.4}_{-0.3}$ & $2.04\times10^{1}$ & $-$  \\
 J1611$-$5209 & $9.6^{+20.0}_{-19.2}$ & $-19.7^{+0.3}_{-0.3}$ & $2.39\times10^{2}$ & $-$  \\
 J1632$-$4757 & $10.3^{+20.1}_{-18.8}$ & $-19.3^{+0.4}_{-0.3}$ & $1.67\times10^{2}$ & $-$  \\
 J1637$-$4553 & $10.4^{+19.7}_{-18.9}$ & $-19.6^{+0.4}_{-0.3}$ & $2.04\times10^{2}$ & $-$  \\
 J1638$-$4417 & $10.0^{+19.7}_{-19.1}$ & $-20.0^{+0.4}_{-0.4}$ & $3.32\times10^{2}$ & $-$  \\
 J1640$-$4715 & $10.2^{+19.6}_{-18.5}$ & $-19.3^{+0.4}_{-0.4}$ & $2.10\times10^{2}$ & $-$  \\
 J1649$-$4653 & $9.89^{+20.20}_{-19.40}$ & $-19.1^{+0.4}_{-0.4}$ & $3.29\times10^{2}$ & $-$  \\
 J1702$-$4306 & $11.3^{+19.4}_{-18.8}$ & $-20.0^{+0.4}_{-0.3}$ & $5.75\times10^{1}$ & $-$  \\
 J1722$-$3712 & $9.81^{+20.00}_{-19.10}$ & $-19.0^{+0.4}_{-0.3}$ & $5.61\times10^{2}$ & $-$  \\
 J1723$-$3659 & $9.99^{+19.80}_{-19.10}$ & $-19.2^{+0.4}_{-0.4}$ & $4.35\times10^{2}$ & $-$  \\
 J1733$-$3716 & $11.3^{+19.0}_{-18.0}$ & $-20.4^{+0.4}_{-0.3}$ & $3.59\times10^{1}$ & $-$  \\
 J1735$-$3258 & $9.84^{+20.00}_{-19.00}$ & $-18.9^{+0.4}_{-0.4}$ & $5.84\times10^{2}$ & $-$  \\
 J1739$-$2903 & $10.6^{+18.5}_{-17.7}$ & $-20.9^{+0.4}_{-0.4}$ & $4.02\times10^{1}$ & $-$  \\
 J1739$-$3023 & $7.92^{+16.20}_{-15.90}$ & $-19.5^{+0.4}_{-0.3}$ & $1.95\times10^{1}$ & $-$  \\
 J1745$-$3040 & $8.48^{+18.40}_{-18.10}$ & $-20.7^{+0.4}_{-0.4}$ & $4.38\times10^{1}$ & $-$  \\
 J1801$-$2154 & $10.3^{+20.1}_{-18.9}$ & $-19.6^{+0.3}_{-0.3}$ & $3.36\times10^{2}$ & $-$  \\
 J1820$-$1529 & $10.2^{+19.7}_{-18.9}$ & $-18.9^{+0.4}_{-0.3}$ & $1.95\times10^{2}$ & $-$  \\
 J1825$-$1446 & $11.0^{+19.5}_{-18.8}$ & $-19.6^{+0.4}_{-0.3}$ & $6.52\times10^{1}$ & $-$  \\
 J1828$-$1101 & $10.1^{+18.4}_{-17.6}$ & $-18.4^{+0.4}_{-0.4}$ & $3.86\times10^{1}$ & $-$  \\
 J1832$-$0827 & $4.53^{+14.40}_{-12.20}$ & $-20.5^{+0.3}_{-0.3}$ & $1.07\times10^{1}$ & $-$  \\
 J1834$-$0731 & $9.25^{+18.90}_{-18.40}$ & $-19.8^{+0.4}_{-0.4}$ & $4.46\times10^{1}$ & $-$  \\
 J1835$-$1106 & $8.93^{+19.40}_{-19.20}$ & $-18.7^{+0.4}_{-0.3}$ & $1.17\times10^{2}$ & $-$  \\
 J1837$-$0559 & $9.74^{+19.70}_{-18.80}$ & $-19.8^{+0.4}_{-0.4}$ & $7.23\times10^{2}$ & $-$  \\
 J1838$-$0453 & $9.1^{+19.6}_{-20.1}$ & $-18.3^{+0.3}_{-0.3}$ & $1.56\times10^{2}$ & $-$  \\
 J1839$-$0321 & $10.8^{+20.3}_{-18.9}$ & $-19.7^{+0.3}_{-0.3}$ & $1.25\times10^{2}$ & $-$  \\
 J1839$-$0905 & $11.0^{+20.4}_{-19.2}$ & $-19.6^{+0.3}_{-0.2}$ & $2.31\times10^{2}$ & $-$  \\
 J1842$-$0905 & $9.67^{+19.60}_{-19.20}$ & $-19.8^{+0.4}_{-0.3}$ & $2.97\times10^{2}$ & $-$  \\
 J1843$-$0702 & $9.81^{+20.00}_{-19.00}$ & $-19.9^{+0.4}_{-0.4}$ & $1.09\times10^{3}$ & $-$  \\
 J1844$-$0538 & $11.3^{+20.3}_{-18.4}$ & $-20.2^{+0.3}_{-0.3}$ & $6.55\times10^{1}$ & $-$  \\
 J1853$-$0004 & $9.92^{+19.60}_{-19.50}$ & $-18.6^{+0.4}_{-0.4}$ & $4.96\times10^{2}$ & $-$  \\
\hline
\end{tabular*}
\end{center}
\end{table*}

\section{Population trends} \label{Sec:Discussion}

The per-pulsar inference results in Table~\ref{Tab:per-pulsar_inf} exhibit interesting trends across the sample. In Section~\ref{subsecV:PPdot}, we study how the inferred $(n_{\rm pl}+\dot{K}_{\rm dim})^{(m)}$ and $\chi^{(m)}$ values vary across a traditional $P$-$\dot{P}$ diagram, with $P=\nu^{-1}$ and $\dot{P} = -\nu^{-2}\dot{\nu}$.  In Section~\ref{subsecV:correlations}, we study how the spin-down first derivative $\dot{\nu}$ and hence $\tau_{\rm sd} \propto \nu / \dot{\nu}$ correlate with $\ddot{\nu}$ before and after correcting for the stochastic braking index anomaly caused by $\sigma_{\ddot{\nu}} \neq 0$. In Section~\ref{subsecV:comparisonAD}, we compare the results in this paper with those from the previous analyses done by~\cite{ParthasarathyShannon2019} and \cite{ParthasarathyJohnston2020}.

\subsection{$P$-$\dot{P}$ plane} \label{subsecV:PPdot}

The location of pulsars in the $P$-$\dot{P}$ plane is a traditional diagnostic tool in pulsar astronomy. In particular, how pulsars move in the $P$-$\dot{P}$ plane communicates information about their braking torques~\citep{JohnstonKarastergiou2017}. 

\begin{figure*}
\begin{center}
 \includegraphics[width=\textwidth]{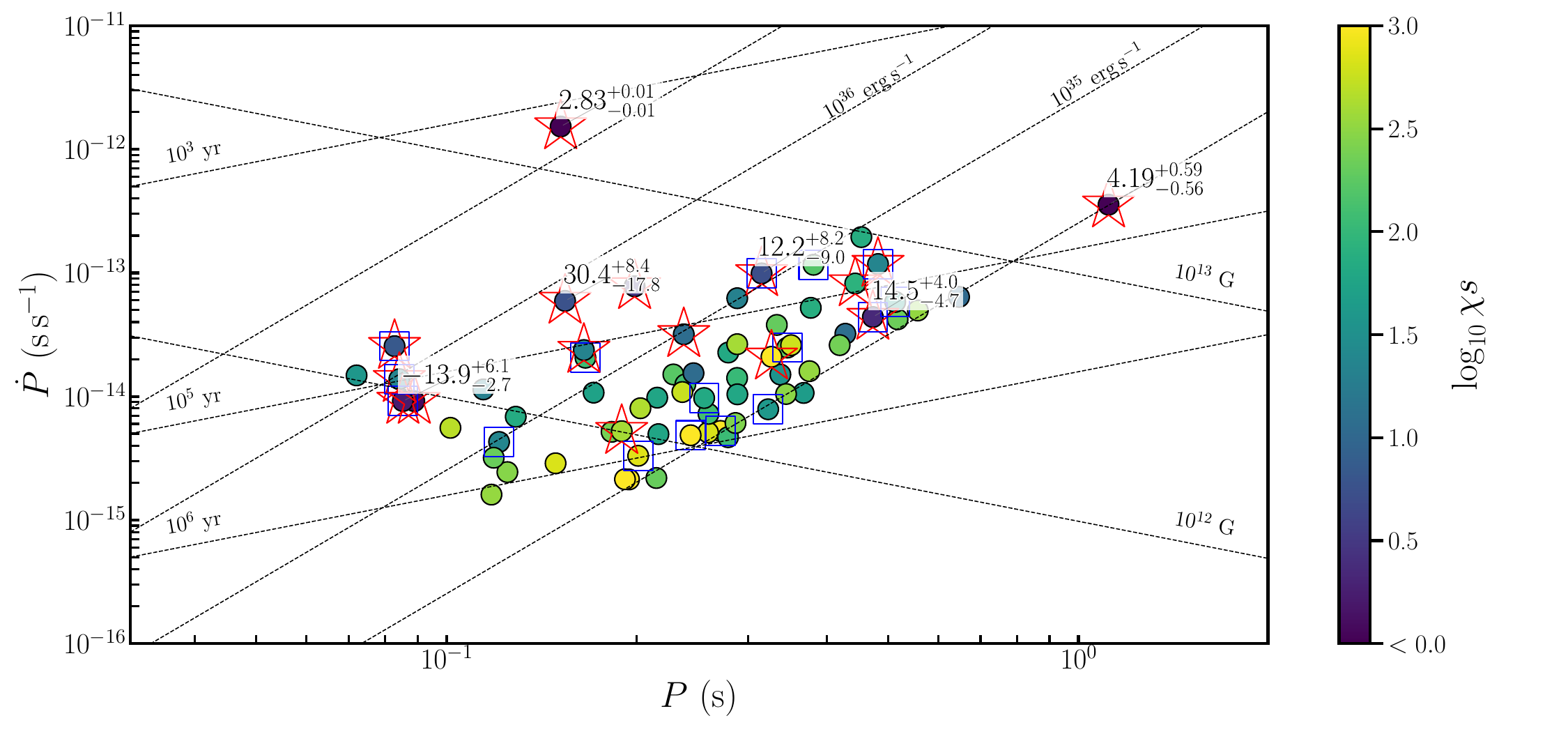}
 \caption{Location of the $68$ P574 pulsars analyzed in this paper in the plane spanned by period $P=\nu^{-1}$ (horizontal axis; logarithmic scale) versus spin-down rate $\dot{P}=-\nu^{-2}\dot{\nu}$ (vertical axis; logarithmic scale). Dashed, black lines indicate constant characteristic age $\tau_{\rm sd}$, surface magnetic field $B$, and spin-down luminosity $\dot{E}$, in units of ${\rm yr}$, ${\rm G}$, and ${\rm erg\,s}^{-1}$ respectively. The color scheme indicates the inferred  stochastic braking index anomaly $\chi s$ (logarithmic scale). A blue square decorating a data point indicates a pulsar that has glitched at least once, while a red star indicates a pulsar with a previously published measurement of $n$ from~\protect \cite{ParthasarathyJohnston2020}. We report the inferred posterior median $n_{\rm pl}+\dot{K}_{\rm dim}$  and $90\%$ credible intervals, alongside a few representative objects which satisfy $n_{\rm pl}+\dot{K}_{\rm dim} \geq \chi s$ i.e. for which the stochastic anomaly is smaller than the secular anomaly. 
 }
\label{fig_secV:ppdot}
\end{center}
\end{figure*}

Fig.~\ref{fig_secV:ppdot} locates the 68 pulsars analyzed in this paper as points in the $P$-$\dot{P}$ plane. The color of each point indicates the magnitude of $\chi s$, the square root of the second term on the right-hand side of equation~(\ref{Eq_SecI:Variance_n}), i.e. the characteristic magnitude of the stochastic braking index anomaly caused by $\sigma_{\ddot{\nu}} \neq 0$. Dark points indicate that the measured braking index is dominated by secular braking with $\dot{K}=0$ or $\dot{K} \neq 0$. Light points indicate that the measured braking index is dominated by $\ddot{\nu}$ fluctuations associated with the mean-reverting random walk described in Section~\ref{subsecII:BrownianModel}. Numbers indicate the median value of $n_{\rm pl} +\dot{K}_{\rm dim}$ inferred by the hierarchical Bayesian scheme in Section~\ref{Sec:Analysis}, for objects where the secular anomaly dominates the stochastic anomaly, i.e.\ $\chi^{(m)} s^{(m)} \leq (n_{\rm pl} + \dot{K}_{\rm dim})^{(m)}$. The red markers decorating some points represent a pulsar that has glitched at least once (square) and/or possesses a previously published measurement of $n$ \citep{ParthasarathyJohnston2020} (star).\footnote{We count a pulsar as having glitched, if it appears in the Jodrell Bank glitch catalogue (see footnote~2) or the Australian Telescope National Facility (ATNF) glitch catalogue: \href{https://www.atnf.csiro.au/research/pulsar/psrcat/glitchTbl.html}{https://www.atnf.csiro.au/research/pulsar/psrcat/glitchTbl.html}} 

We observe the following trends in Fig.~\ref{fig_secV:ppdot}.

\begin{enumerate}
    \item Two objects exhibit $(n_{\rm pl}+\dot{K}_{\rm dim})$-posteriors satisfying $2 \leq n_{\rm pl}+\dot{K}_{\rm dim} \leq 7$, at $90\%$ confidence, with $\chi s \leq 0.5$. They are PSR J1513$-$5908 and PSR J1524$-$5704. Both objects are young, with $\tau_{\rm sd} \leq 5 \times10^{4} ~{\rm yr}$, and possess the highest (canonical) surface magnetic field strengths in the sample, with $B \gtrsim 1.5\times10^{13}~{\rm G}$. That is, they reside near the top of the $P$-$\dot{P}$ diagram in Fig.~\ref{fig_secV:ppdot}. Other objects studied elsewhere with nonanomalous measured braking indices $1 \leq n \leq 3.15$ are also young and highly magnetized~\citep{LivingstoneKaspi2007,WeltevredeJohnston2011,LyneJordan2015,ArchibaldGotthelf2016,EspinozaLyne2017,Espinoza2018,AkbalGugercinoglu2021}. 
    \item Objects whose measured braking indices are dominated by stochastic [second term on the right-hand side of equation~(\ref{Eq_SecI:Variance_n})] rather than secular [first term on the right-hand side of equation~(\ref{Eq_SecI:Variance_n})] anomalies are in the majority. Among the $68$ pulsars, $58$ have $\chi s \geq |n_{\rm pl} + \dot{K}_{\rm dim} |$. They mainly lie in the core of the $P$-$\dot{P}$ plane bracketed by spin-down luminosities $1 \lesssim \dot{E}/(10^{34}~{\rm erg}\,{\rm s}^{-1}) \lesssim 10$. The stochastic anomaly dominates the secular anomaly increasingly, as the objects age, i.e.\ as one moves from the top left to the bottom right of the $P$-$\dot{P}$ diagram in Fig.~\ref{fig_secV:ppdot}. For example, PSR J0857$-$4424 is an older object, which has $\tau_{\rm sd}=247.48\times10^{3}~{\rm years}$, $\chi s=1.18\times10^{3}$, and a reported braking index of $n=2890^{+30}_{-30}$~\citep{ParthasarathyJohnston2020}.

    \item The inferred median $\chi$-values are related to a pulsar's location in the $P$-$\dot{P}$ plane. A non-linear least squares regression yields

    \begin{equation}
        \chi = 10^{-12.33\pm1.97} \Big(\frac{P}{1~{\rm s}}\Big)^{-1.21\pm0.35}\Big(\frac{\dot{P}}{1~{\rm s}\,{\rm s}^{-1}}\Big)^{0.59\pm0.15}~{\rm s}^{-5/2}. \label{eq_secV:chiPPdot}
    \end{equation}
    
    The reduced chi-square statistic for the fit is $0.89$. Equation~(\ref{eq_secV:chiPPdot}) can be expressed equivalently as $\chi \propto \nu^{0.038\pm0.34}\vert \dot{\nu} \vert^{0.59\pm0.15}$, nearly independent of $\nu$. Physically, this makes sense, if $\sigma_{\ddot{\nu}}$ is caused by internal or magnetospheric fluctuations driven by spin down, e.g. due to differential rotation  between the crust and superfluid core~\citep{MelatosLink2014}. We note that equation~(\ref{eq_secV:chiPPdot}) is consistent with trend (ii) above.
    
    \item In nine objects, the $(n_{\rm pl}+\dot{K}_{\rm dim})$-posteriors exclude zero at $90\%$ confidence. All of them satisfy $\chi s \leq 7$. All nine objects are scattered, in no apparent pattern, throughout the $P$-$\dot{P}$ plane. Four have glitched previously. Among the latter, PSR J1833$-$0827 returns $n_{\rm pl} +\dot{K}_{\rm dim} < 0$, the only object out of the 68 to do so. This is consistent with previous studies, which find that positive anomalous braking indices are more common than negative anomalous braking indices~\citep{ChukwudeBaiden2010, ParthasarathyJohnston2020, LowerBailes2020}. However, the imbalance is more lopsided in this paper, viz.\ one object out of 68, as compared to two objects out of 19 in~\cite{ParthasarathyJohnston2020} or four objects out of nine in~\cite{LowerBailes2020}.
\end{enumerate}

A $P$-$\dot{P}$ scaling for $n_{\rm pl}+\dot{K}_{\rm dim}$ can be fitted empirically in principle, analogously to Equation~(\ref{eq_secV:chiPPdot}). The result is the same as the standard scaling quoted previously in the literature~\citep{CordesChernoff1998,Faucher-GiguereKaspi2006} and is not repeated here for the sake of brevity.

\subsection{Correlations between first and second $\nu$ derivatives} \label{subsecV:correlations}

\begin{figure}
\begin{center}
 \includegraphics[width=\columnwidth]{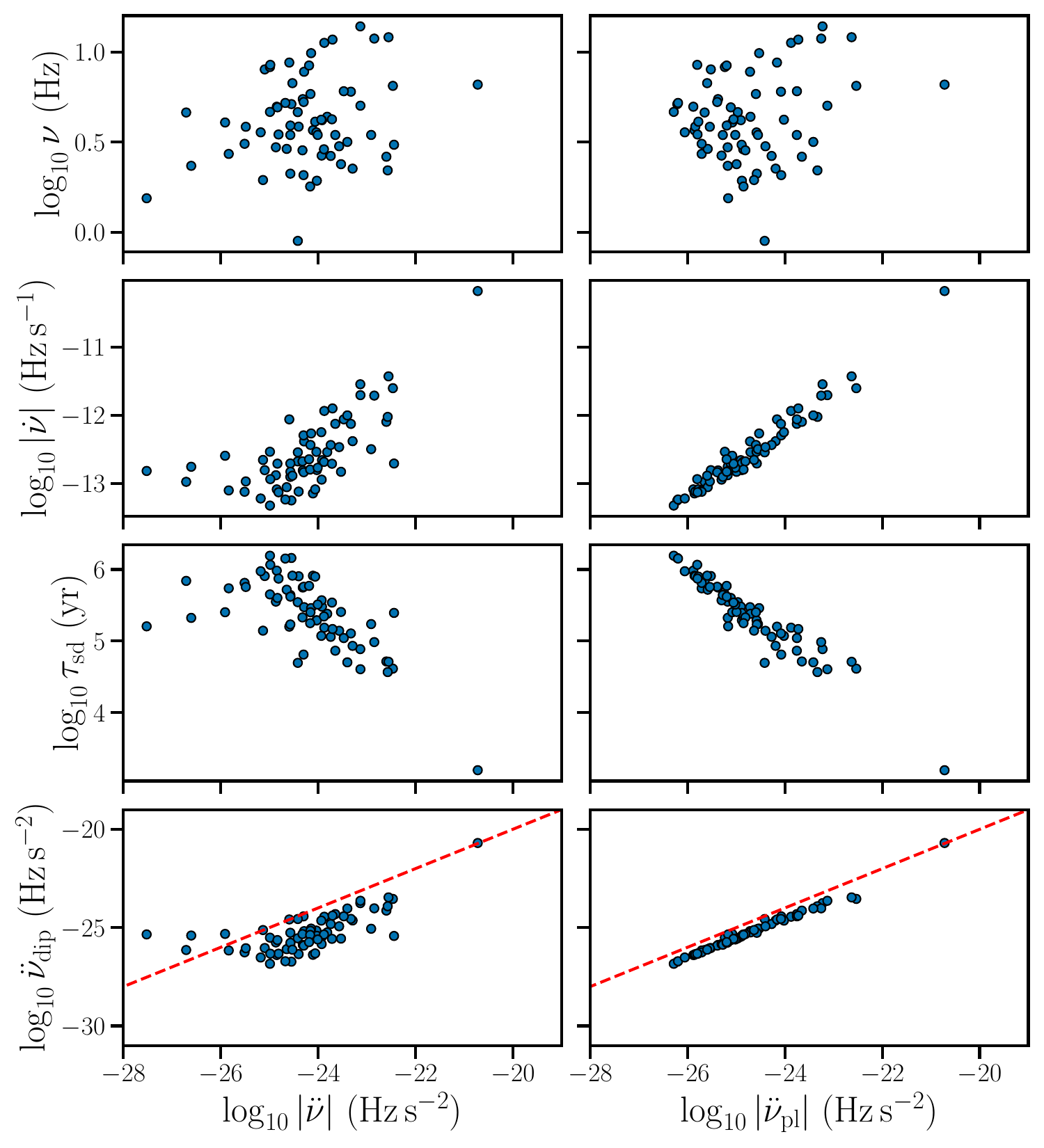}
 \caption{Cross-correlations between the rotational parameters $\nu,\dot{\nu}, \tau_{\rm sd}$, and $\ddot{\nu}_{\rm dip}$ (top to bottom rows respectively) and the second frequency derivative  measured by~\tempoDOS~($\vert \ddot{\nu} \vert$; left column) and the hierarchical Bayesian scheme~($\vert \ddot{\nu}_{\rm pl}\vert$; right column). In the last row, the red, dashed line indicates $\ddot{\nu}_{\rm dip}=\vert \ddot{\nu} \vert$ and $\ddot{\nu}_{\rm dip}=\vert \ddot{\nu}_{\rm pl} \vert$. PSR J1513$-$5908, with $n_{\rm pl}+\dot{K}_{\rm dim}=2.83^{+0.01}_{-0.01}$, is the closest object to the red, dashed line.}
\label{fig_secV:correlations}
\end{center}
\end{figure}

\begin{table}
\begin{center}
\caption{Cross-correlation coefficients between $\vert \ddot{\nu} \vert$ and the parameters $\nu,\dot{\nu}$, $ \tau_{\rm sd}=\nu/(2\vert \dot{\nu} \vert)$, and $\ddot{\nu}_{\rm dip}=\ddot{\nu}_{n_{\rm pl}=3, \dot{K}_{\rm dim}=0}=3\dot{\nu}^{2}\nu^{-1}$. In the top row of the table, we use the value of $\ddot{\nu}$ obtained from \tempoDOS. In the bottom row, we use the inferred, median values $\ddot{\nu}_{\rm pl}=(n_{\rm pl}+\dot{K}_{\rm dim})\dot{\nu}^{2}\nu^{-1}$ from the Bayesian scheme. The error bars give the standard error of each Pearson coefficient~\citep{BoslaughWatters2008}. } 
\label{Tab:correlations}
\begin{tabular*}{\columnwidth}{@{\extracolsep{\fill}}ccccc}
\hline
   \; & $\nu$ & $\vert\dot{\nu}\vert$ & $\tau_{\rm sd}$ & $\ddot{\nu}_{\rm dip}$ \\
   \hline
   $\vert \ddot{\nu} \vert$ & $0.23\pm0.12$ & $0.69\pm0.06$ & $-0.64\pm0.07$ & $0.68\pm0.07$ \\
   $\vert \ddot{\nu}_{\rm pl} \vert$ &  $0.21\pm0.12$ & $0.98\pm0.01$ & $-0.95\pm0.01$ & $0.99\pm0.01$ \\
  \hline 
\end{tabular*}
\end{center}
\end{table}

Observational evidence exists in support of correlations between timing noise amplitude and spin-down derivatives, such as $\dot{\nu}$, and hence $\tau_{\rm sd}$~\citep{CordesHelfand1980, ArzoumanianNice1994, BeskinBiryukov2006,UramaLink2006}. The amplitude of fluctuations in $\ddot{\nu}$ is taken sometimes as a proxy for the timing noise amplitude. For example, one possible explanation for the observed $|\ddot{\nu}|$-$\dot{\nu}$ trend is that a stochastic torque drives fluctuations around the secular braking torque~\citep{UramaLink2006, VargasMelatos2023}, i.e. scenario (ii) in Sections~\ref{Sec:Introduction} and~\ref{subsecII:BrownianModel}. Often, however, the stochastic and secular contributions to $| \ddot{\nu} |$ are comparable in amplitude, so it is unclear which of the two contributions is mainly responsible for the observed $| \ddot{\nu} |$-$\dot{\nu}$ trend, or if both contribute similarly. We study this issue in this section.

Leveraging the ability of the hierarchical Bayesian scheme in Section~\ref{subsecII:Bayesframework} to disentangle the stochastic and secular contributions at the per-pulsar and population levels, one can ask how the correlation between $\vert \ddot{\nu}\vert$ and $\dot{\nu}$ changes, when we use the inferred, median, per-pulsar values $\ddot{\nu}_{\rm pl}^{(m)}=(n_{\rm pl}+\dot{K}_{\rm dim})^{(m)}[\dot{\nu}^{(m)}]^{2}[\nu^{(m)}]^{-1}$, i.e. the secular second frequency derivative excluding stochastic fluctuations. The goal is to test whether the correlation tightens or loosens, when we use $\ddot{\nu}_{\rm pl}^{(m)}$ instead of the timing-noise-contaminated measurement $\ddot{\nu}$. For the sake of completeness, we follow~\cite{UramaLink2006} and cross-correlate $\vert \ddot{\nu}_{\rm pl} |$ and $|\ddot{\nu}|$ against not only $\dot{\nu}$ but also the related first-derivative quantities, $\tau_{\rm sd} = \nu / (2\dot{\nu})$, and $\ddot{\nu}_{\rm dip}=\ddot{\nu}_{n_{\rm pl}=3, \dot{K}_{\rm dim}=0}=3\dot{\nu}^{2}\nu^{-1} \neq \ddot{\nu}_{\rm pl}$. The latter quantities assume implicitly that the secular torque is dominated by magnetic dipole braking ($n_{\rm pl}=3$) with $\dot{K}_{\rm dim}=0$. In reality, the evidence in favor of $\dot{K}_{\rm dim} \neq 0$ is strong, as noted in Sections~\ref{Sec:Analysis}--\ref{Sec:astro_int}.

Figure~\ref{fig_secV:correlations} graphs the trends between the second frequency derivatives $| \ddot{\nu} |$ and $| \ddot{\nu}_{\rm pl} |$ and $\nu$ (first row), $\vert \dot{\nu} \vert$ (second row), $\tau_{\rm sd}$ (third row), and $\ddot{\nu}_{\rm dip}$ (fourth row) for the 68 P574 pulsars analyzed in this paper. In the left column, $ | \ddot{\nu} |$ is calculated in the traditional manner by~\tempoDOS. In the right column, $\ddot{\nu}_{\rm pl} = (n_{\rm pl}+\dot{K}_{\rm dim})^{(m)}[\dot{\nu}^{(m)}]^{2}[\nu^{(m)}]^{-1}$ is calculated from the median estimate of the hierarchical Bayesian scheme in Table~\ref{Tab:per-pulsar_inf}, with the stochastic anomaly subtracted. Table~\ref{Tab:correlations} summarizes the associated Pearson correlation coefficients, with $\vert \ddot{\nu}\vert$ (left column of Fig.~\ref{fig_secV:correlations}) and $\vert \ddot{\nu}_{\rm pl}\vert$ (right column of Fig.~\ref{fig_secV:correlations}) corresponding to the top and bottom rows of the table respectively. It is striking that the first-derivative correlations are clearly tighter with the median $| \ddot{\nu}_{\rm pl} |$ than with $| \ddot{\nu} |$. This is expected; the confounding influence of the stochastic anomaly is present in $| \ddot{\nu} |$ but absent from $| \ddot{\nu}_{\rm pl} |$. When using $\vert\ddot{\nu}\vert$, the strongest correlations are $\vert \dot{\nu} \vert, \ddot{\nu}_{\rm dip},$ and $\tau_{\rm sd}$ (negative) in descending order . In contrast, when using $\vert \ddot{\nu}_{\rm pl} \vert$, the strongest correlations are $\ddot{\nu}_{\rm dip}$ with $99\%$, $\vert \dot{\nu} \vert$ with $98\%$, and $\tau_{\rm sd}$ with $95\%$ (negative). We note that the correlations $\vert \ddot{\nu}_{\rm pl}\vert$-$\vert \dot{\nu}\vert$, $\vert \ddot{\nu}_{\rm pl}\vert$-$\tau_{\rm sd}$, and $\vert \ddot{\nu}_{\rm pl}\vert$-$\ddot{\nu}_{\rm dip}$, span $5.6~{\rm dex}$ of $\vert \ddot{\nu}_{\rm pl}\vert$ values.

Is it possible to exploit the correlation between $\ddot{\nu}_{\rm pl}$ and $\ddot{\nu}_{\rm dip}$ to learn more about $\dot{K}_{\rm dim} \propto \tau_{\rm sd} / \tau_K$? Yes, up to a point. It is impossible to measure $\dot{K}_{\rm dim}$ directly, because it always appears degenerately with $n_{\rm pl}$ in the sum $n_{\rm pl} + \dot{K}_{\rm dim}$. Stated otherwise, $\ddot{\nu}_{\rm pl}$ correlates with $n_{\rm pl} \dot{\nu}^2 \nu^{-1}$ equally for all $n_{\rm pl}$, including $2\lesssim n_{\rm pl} \lesssim 7$ for standard mechanisms (see Section~\ref{Sec:Introduction}). However, if one assumes some physical knowledge of $n_{\rm pl}$ a priori, one can make progress. For example, one may assume $n_{\rm pl}=3$, fit the median values $\ddot{\nu}_{\rm pl}$ with $\log_{10} \vert \ddot{\nu}_{\rm pl} \vert = m\log_{10} \vert \ddot{\nu}_{\rm dip} \vert+b$ using nonlinear least squares, and find $m = 0.98\pm0.02$ and $b=0.25\pm0.44$.  From the fit we infer two population-wide bounds on $\dot{K}_{\rm dim}$. First, ${\rm mean}\{p[(n_{\rm pl}+\dot{K}_{\rm dim})^{(\rm pop)}\vert D]\}>3$ and $\dot{K}_{\rm dim} = (\dot{K}/K)(\nu/\dot{\nu})$ imply $\dot{K}/K < 0$ on average for the population. Second, $n=n_{\rm pl}+K_{\rm dim}$ and  $K_{\rm dim} = (\dot{K}/K)(\nu/\dot{\nu})$ imply that the least-squares intercept $b$ satisfies $\vert\dot{K}/K\vert\tau_{\rm sd}= n_{\rm pl}(n_{\rm pl}-1)^{-1}(10^{b}-1)$. The above regression yields a central value of $\vert \dot{K}/K \vert\tau_{\rm sd} = 1.16$, with one-sigma range $0 \leq \vert \dot{K}/K \vert\tau_{\rm sd} \leq 5.8$. That is, for the 68 P574 pulsars,  $\tau_{K} \sim \vert K/\dot{K}\vert$ is on average comparable to $\tau_{\rm sd}$. The same holds for six integer values of $n_{\rm pl}$ in the range $2\leq n_{\rm pl} \leq 7$. The result $\tau_{K} \sim \tau_{\rm sd}$ is interpreted physically in Section~\ref{Sec:astro_int}. 

There are limitations to the approach in the previous paragraph. In particular, the estimates parameters $m$ and $b$ do not take into account the error bars on the $\ddot{\nu}_{\rm pl}^{(m)}$ estimates. If instead the fit $\log_{10} \vert \ddot{\nu}_{\rm pl} \vert = m\log_{10} \vert \ddot{\nu}_{\rm dip} \vert+b$ is weighted by the uncertainties on $\ddot{\nu}_{\rm pl}^{(m)}$, i.e. $[\Delta(n_{\rm pl}+\dot{K}_{\rm dim})^{(m)}][\dot{\nu}^{(m)}]^{2}[\nu^{(m)}]^{-1}$, where $\Delta(n_{\rm pl}+\dot{K}_{\rm dim})^{(m)}$ is the $90\%$ credible interval on $(n_{\rm pl}+\dot{K}_{\rm dim})^{(m)}$ (see Table~\ref{Tab:per-pulsar_inf}), then one finds $m=0.98\pm2.88$ and $b=0.25\pm72.68$. This is expected; the error bars on the per-pulsar estimates for $(n_{\rm pl}+\dot{K}_{\rm dim})^{(m)}$ are broad for the objects satisfying $ \chi s \gg \vert n_{\rm pl}+\dot{K}_{\rm dim}\vert $. Future analyses using larger pulsar samples, which yield more accurate inference results, will improve the constrains on $m$ and $b$.

\subsection{Previous $n$ measurements} \label{subsecV:comparisonAD}

In this section, we compare the braking indices measured by~\cite{ParthasarathyJohnston2020} with those inferred in Section~\ref{Sec:Analysis}. The goal is to understand under what circumstances the hierarchical scheme, which disentangles secular and stochastic contributions to $n$, agrees with traditional methods such as~\temponest~\citep{LentatiAlexander2014,ParthasarathyShannon2019,LowerBailes2020}.

The rightmost column of Table~\ref{Tab:per-pulsar_inf} lists the $n$ values measured by \citet{ParthasarathyJohnston2020} for 16 P574 pulsars (top half of Table~\ref{Tab:per-pulsar_inf}). We observe that $n_{\rm pl} + \dot{K}_{\rm dim}$, as inferred by the hierarchical Bayesian scheme (second column), and $n$, as measured by \cite{ParthasarathyJohnston2020} (fifth column), are consistent for $\chi s \lesssim 10$. For example, for PSR J1809$-$1917,  we infer $\chi s = 6.77$, and $n=23.5^{+6.0}_{-6.0}$ is bracketed by $n_{\rm pl}+\dot{K}_{\rm dim}=20.35^{+9.33}_{-13.48}$. In contrast, for PSR J1648$-$4611, we infer $\chi s = 21.7$, and $n=40.0^{+10.0}_{-10.0}$ barely overlaps the upper boundary of $n_{\rm pl}+\dot{K}_{\rm dim}=14.04^{+17.40}_{-17.14}$. For PSR J0857$-$4424, which has the highest $\chi s$ value in the top half of Table~\ref{Tab:per-pulsar_inf}, we infer $n_{\rm pl}+\dot{K}_{\rm dim} \ll n$. The empirical condition $\chi s \lesssim 10$ is a rough visual proxy for the physically motivated condition $\chi s \leq | n_{\rm pl} + \dot{K}_{\rm dim}|$ identified in point (ii) in Section~\ref{subsecV:PPdot}. That is, the $n$ value measured by \cite{ParthasarathyJohnston2020} and the $n_{\rm pl} + \dot{K}_{\rm dim}$ value inferred by the hierarchical Bayesian scheme are consistent, when the secular contribution to $\ddot{\nu}$ exceeds the stochastic contribution.

\begin{figure}
\flushleft
 \includegraphics[width=\columnwidth]{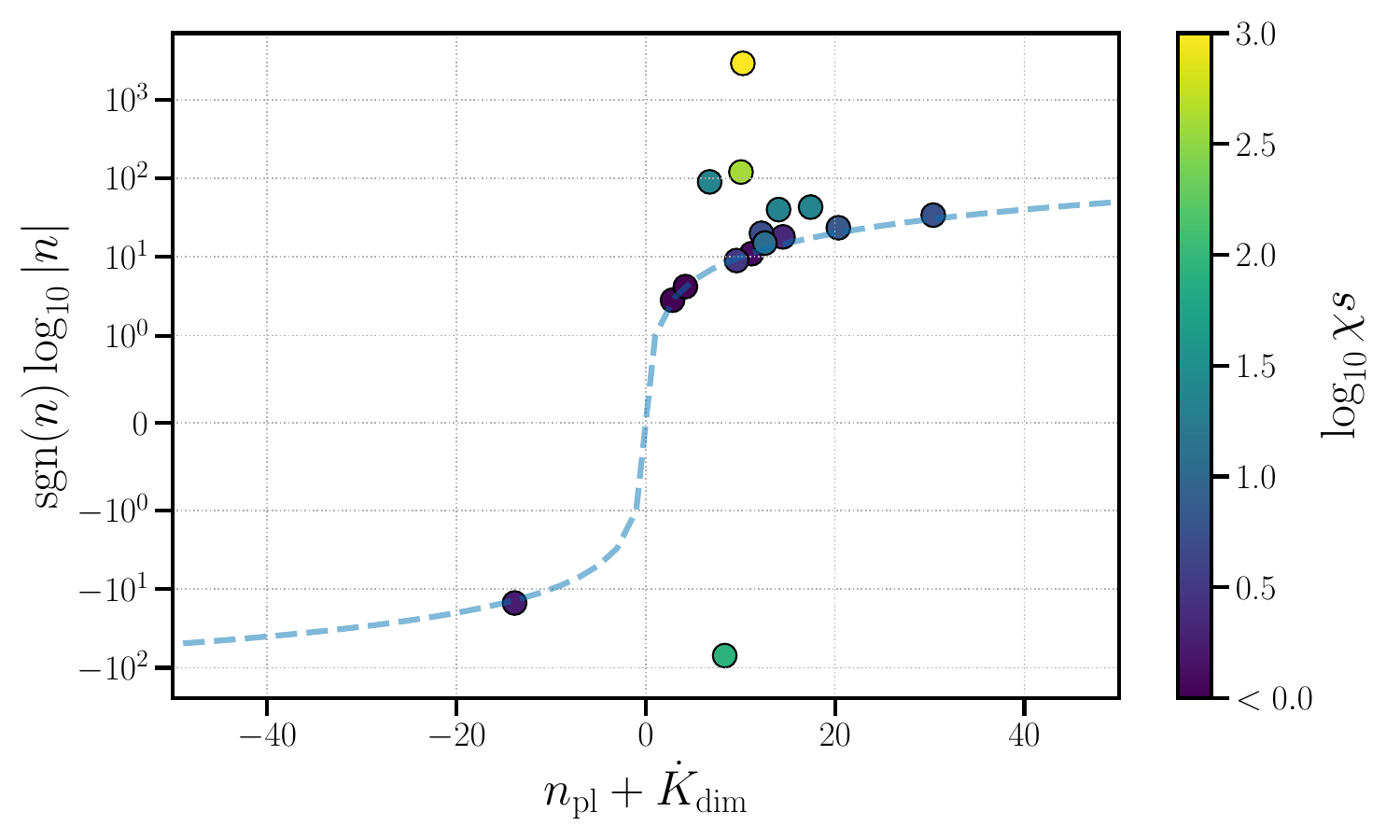}
 \caption{Correcting for the stochastic anomaly in braking index measurements: values of $n_{\rm pl}+\dot{K}_{\rm dim}$ inferred by the hierarchical Bayesian scheme, which exclude the stochastic anomaly  (horizontal axis; see Section~\ref{subsecII:Bayesframework}), versus traditional~\temponest~measurements of $n$, reported by \protect \cite{ParthasarathyJohnston2020} (vertical axis; logarithmic scale). The blue, dashed curve indicates $n=n_{\rm pl}+\dot{K}_{\rm dim}$. The color scale indicates $\chi s$, the root-mean-square amplitude of the stochastic anomaly [see equation~(\ref{Eq_SecI:Variance_n})]. All pulsars satisfying $\chi s \leq 1$ are assigned the color at the lower end of the scale. Objects that lie further from the dashed curve correspond to higher $\chi s$ values, i.e. their stochastic anomaly dominates their secular anomaly~\protect \citepalias{VargasMelatos2024}.}
\label{fig_secV:comparison_nTN_vs_nBM}
\end{figure}

Fig.~\ref{fig_secV:comparison_nTN_vs_nBM} offers a visual representation of the property discussed in the previous paragraph. We plot the inferred $n_{\rm pl}+\dot{K}_{\rm dim}$ values (horizontal axis) versus the measured $n$ (signed, logarithmic scale; vertical axis), while the color scale represents $\log_{10}\chi s$. For higher values of $\log_{10} \chi s$, the objects lie above the blue, dashed curve $n=n_{\rm pl}+\dot{K}_{\rm dim}$, because the stochastic anomaly $\chi s$ is additive [see Equation~(\ref{Eq_SecI:Variance_n})].

\section{Astrophysical Interpretation} \label{Sec:astro_int}

The hyperparameters $\mu_{\rm pl} = 9.95^{+5.58}_{-5.26}$ and $\sigma_{\rm pl} = 10.89^{+5.14}_{-3.69}$ inferred in Section~\ref{Sec:Analysis} cover an astrophysically interesting domain. The associated posterior distribution $p[(n_{\rm pl}+\dot{K}_{\rm dim})^{(\rm pop)} \vert D]$ overlaps with the canonical power-law braking indices associated with magnetic dipole braking $(n_{\rm pl}=3)$~\citep{GunnOstriker1969} and mass or current quadrupole gravitational radiation reaction ($n_{\rm pl}=5$ or $7$)~\citep{Thorne1980}, with $13\%$ of the posterior mass lying within $3 \leq n_{\rm pl} + \dot{K}_{\rm dim} \leq 7$. However, the population-level prior admits a wide range of $n_{\rm pl} + \dot{K}_{\rm dim}$ values outside the canonical range, with $26\%$ and $61\%$ of the posterior mass lying at $n_{\rm pl} + \dot{K}_{\rm dim} < 3$ and $n_{\rm pl} + \dot{K}_{\rm dim} > 7$, respectively. The results in Section~\ref{Sec:Analysis} are also consistent with independent analyses, that do not rely on subtracting the timing noise contribution from anomalous braking indices. For example, population synthesis calculations based on proper motion measurements of young pulsars, without reference to $\ddot{\nu}$ data, favor $n_{\rm pl} \approx 4.5\pm0.5$, if $K$ is constant, and $n_{\rm pl}\approx 2.5$, if $K$ evolves~\citep{CordesChernoff1998, Faucher-GiguereKaspi2006}. In this section, we discuss three possible astrophysical interpretations of the results $\mu_{\rm pl} = 9.95^{+5.58}_{-5.26}$ and $\sigma_{\rm pl} = 10.89^{+5.14}_{-3.69}$, namely a single secular braking mechanism with $3 \lesssim n_{\rm pl} \lesssim 7$ and $\dot{K}_{\rm dim} = 0$~(Section~\ref{subsecVI:single_pl_with_no_Kdot}), two mechanisms with $3 \lesssim n_{\rm pl} \lesssim 7$ and $\dot{K}_{\rm dim} =0$ (Section~\ref{subsecVI:two_pl_with_no_Kdot}), and a single mechanism with $3 \lesssim n_{\rm pl} \lesssim 7$ and $\dot{K}_{\rm dim} \neq 0$~(Section~\ref{subsecVI:single_pl_with_Kdot}). The three interpretations are not exhaustive; many other valid scenarios exist. It is difficult to  count reliably the number (one, two, or more) of distinct secular mechanisms, when one has $n_{\rm pl} \lesssim \vert \dot{K}_{\rm dim} \vert$, and $n_{\rm pl} + \dot{K}_{\rm dim}$ overlaps (due to $\dot{K}_{\rm dim}\neq 0$) for different mechanisms.

\subsection{Single power-law mechanism with $\dot{K}_{\rm dim} = 0$} \label{subsecVI:single_pl_with_no_Kdot}

A single secular braking mechanism with $\dot{K} = 0$ is consistent with the posterior $p(\mu_{\rm pl}, \sigma_{\rm pl}\vert D)$ calculated in Section~\ref{Sec:Analysis}. The posterior is unimodal and has negligible probability weight near the left and right edges of the prior domain $-75 \leq n_{\rm pl}+\dot{K}_{\rm dim} \leq 75$. Note that a single secular braking mechanism with $\dot{K}_{\rm dim}=0$ does not imply necessarily that $n_{\rm pl} + \dot{K}_{\rm dim} = n_{\rm pl}$ is the same in every pulsar. This is exemplified by the hierarchical Bayesian results in Table~\ref{Tab:per-pulsar_inf}, where for example $n_{\rm pl} + \dot{K}_{\rm dim}$ does not overlap for PSR J1513$-$5908 and PSR J1524$-$5706 at $90\%$ confidence, as well as by previous measurements of nonanomalous braking indices in the literature, which return nonoverlapping values $2 \lesssim n_{\rm pl} \lesssim 3$ consistent with a single mechanism (viz.\ magnetic dipole braking)~\citep{LivingstoneKaspi2007,LivingstoneKaspi2011,ArchibaldGotthelf2016}

If indeed a single mechanism operates, what is it physically? One possibility is canonical magnetic dipole braking, with $n_{\rm pl} \approx 3$~\citep{GunnOstriker1969}. The $13$ nonanomalous braking indices that have been measured to an accuracy of better than $10\%$ (say) by using a single coherent timing solution span the range $1.0 \leq n_{\rm meas} \leq 3.15$~\citep{LivingstoneKaspi2007,WeltevredeJohnston2011,LyneJordan2015,ArchibaldGotthelf2016,EspinozaLyne2017,Espinoza2018,AkbalGugercinoglu2021}, which encompasses $6\%$ of the total probability weight in $p(\mu_{\rm pl},\sigma_{\rm pl}\vert D)$, and includes one of the $68$ median values of $(n_{\rm pl} + \dot{K}_{\rm dim})^{(m)}$ in Table~\ref{Tab:per-pulsar_inf}. Variants of magnetic dipole braking are also broadly consistent with $p(\mu_{\rm pl},\sigma_{\rm pl}\vert D)$. Electromagnetic torques exerted by a force-free magnetosphere and wind, with a corotating closed zone terminating at an equatorial y-point, can be approximated by a power-law net torque with $n_{\rm pl}\leq3$ for young pulsars and $n_{\rm pl}\geq3$ for pulsars near the death line~\citep{ContopoulosSpitkovsky2006,Petri2021}. Electromagnetic torques acting on an extended magnetic dipole composed of the star and its corotating, conducting magnetosphere lead to $2 \lesssim n_{\rm pl} \lesssim 3$~\citep{Melatos1997}.

One natural way to push $n_{\rm pl}$ significantly above $n_{\rm pl}=3$ while keeping $\dot{K}=0$, to match more closely the peak of the inferred  distribution $p[(n_{\rm pl}+\dot{K}_{\rm dim})^{(m)} \vert D]$, is to introduce magnetic multipoles of quadrupole and higher order~\citep{Petri2015,Petri2017,AraujoDeLorenci2024}. A pure magnetic quadrupole experiences a pure power-law spin-down torque with $n_{\rm pl}=5$, for example. A magnetic quadrupole superposed on a magnetic dipole experiences a torque, which can be approximated as a power law with $3 \leq n_{\rm pl} \leq 5$. Higher-order multipoles of order $l$ have $n_{\rm pl}=2l+1$. The main reservation concerning this scenario is whether or not the formation mechanism of a radio pulsar's magnetic field~\citep{BraithwaiteSpruit2004,BraithwaiteNordlund2006} can channel enough energy into higher-order multipoles to raise $n_{\rm pl}$ significantly above $n_{\rm pl}=3$, especially for slower rotators. To see why, let $B_{d}$ and $B_{q}$ denote the magnetic field strengths of the dipole and quadrupole respectively at the stellar surface (radius $R_{\star}$). The braking torque is proportional to the Poynting flux at the light cylinder [radius $R_{L}=c/(2\pi\nu)$], which itself is proportional to $B_{d}^{2}(R_{\star}/R_{L})^{6}$ and $B_{q}^{2}(R_{\star}/R_{L})^{8}$ for the dipole and quadrupole respectively~\citep{Petri2015,Petri2019}. One requires $B_{q}^{2}(R_{\star}/R_{L})^{8} > B_{d}^{2}(R_{\star}/R_{L})^{6}$ to obtain $n_{\rm pl} \geq 4$. This implies that the energy stored in the quadrupole must exceed substantially the energy stored in the dipole, with $B_{q}^{2}R_{\star}^{3} \gtrsim 2.28\times10^{5}(\nu/10~{\rm Hz})^{-2}(R_{\star}/10~{\rm km})^{-2}B_{d}^{2}R_{\star}^{3}$. Although magnetars are known to possess magnetic fields with strong multipolar components~\citep{TiengoEsposito2013}, it is unclear at the time of writing, whether the latter inequality is consistent with published mechanisms for magnetic field formation in young, radio pulsars~\citep{FerrarioMelatos2015}.

\subsection{Two power-law mechanisms with $\dot{K}_{\rm dim} = 0$} \label{subsecVI:two_pl_with_no_Kdot}

The posterior $p(\mu_{\rm pl}, \sigma_{\rm pl} \vert D)$ calculated in Section~\ref{Sec:Analysis} is a possible outcome, even if radio pulsars in the Milky Way experience two secular braking mechanisms with $K$ constant. The population-level prior $\pi[n_{\rm pl}^{(m)}, \chi^{(m)} \vert \mu_{\rm pl}, \sigma_{\rm pl}]$ in Equation~(\ref{eq_SecII:npl_Kdim_dist}) assumes a unique, universal braking mechanism for every pulsar. However, the data sample is small enough $(M=68)$, and the dispersion of $n_{\rm pl}$ per mechanism (related to $\sigma_{\rm pl}$) is wide enough, that the hierarchical Bayesian scheme in Section~\ref{Sec:Methods} may interpret a mixture of two broad and closely spaced Gaussians as a single Gaussian.

We emphasize that it is prudent to approach inferences about multiple secular braking mechanisms with caution. The P574 sample is too small to be definitive. Ultimately, one should aim to distinguish between one and two mechanisms by repeating the analysis in Section~\ref{Sec:Analysis} for a two-mechanism population-level prior $\pi[n_{\rm pl}^{(m)}, \chi^{(m)} \vert \mu_{\rm pl}, \sigma_{\rm pl}]$ by constructing an alternative hierarchical Bayesian scheme as a mixture model with hyperparameters $\{\mu_{\rm pl}^{(1)},\sigma_{\rm pl}^{(1)}\},\{\mu_{\rm pl}^{(2)},\sigma_{\rm pl}^{(2)}\}$, and mixing fraction $\lambda$. Such an analysis is postponed, until a larger sample of reliable $n_{\rm meas}$ measurements becomes available, to ensure that the results are significant statistically~\citepalias{VargasMelatos2025}. The task becomes even harder, if one has $n_{\rm pl} \lesssim | \dot{K}_{\rm dim} |$, as discussed in the first paragraph of Section~\ref{Sec:astro_int}.

Magnetic dipole braking ($n_{\rm pl}\approx 3$) and gravitational radiation reaction ($n_{\rm pl}\approx5$ or $7$) represent a natural pair of mechanisms, which are consistent superficially with $\mu_{\rm pl} = 9.95^{+5.58}_{-5.26}$ and $\sigma_{\rm pl} = 10.89^{+5.14}_{-3.69}$ and the per-pulsar $(n_{\rm pl}+\dot{K}_{\rm dim})$-posteriors, except for PSR J0954$-$5430, PSR J1509$-$5850, PSR J1637$-$4642, and PSR J1833$-$0827 (see Table~\ref{Tab:per-pulsar_inf}). However, this scenario poses an astrophysical challenge: it is unlikely that gravitational radiation reaction dominates magnetic dipole braking in many (or indeed any) of the young, radio pulsars analyzed in Section~\ref{Sec:Analysis}, if contemporary thinking is accurate about the magnetic dipole moments~\citep{FerrarioMelatos2015,Petri2019} and mass or current quadrupole moments~\citep{AbbottAbe2022,MoralesHorowitz2022,Riles2023,DergachevPapa2023} of isolated neutron stars. To see why, we equate the power emitted in mass quadrupole gravitational radiation to the rotational energy loss rate to obtain the gravitational-radiation-reaction spin-down rate

\begin{equation}
    \dot{\nu}_{\rm gw} = -5.4\times10^{-13} \Bigg( \frac{I_{zz}}{10^{45}~{\rm g\,cm}^{2}} \Bigg) \Bigg(\frac{\epsilon}{10^{-6}}\Bigg)^{2}\Bigg(\frac{\nu}{100~{\rm Hz}}\Bigg)^{5}~{\rm Hz\,s}^{-1},
\end{equation}

where $I_{zz}$ is the $zz$ component of the moment-of-inertia tensor (${\bf \hat{e}}_{z}$ is the rotation axis), $\epsilon = \vert I_{zz}-I_{xx}\vert /I_{zz}$ is the mass ellipticity, and we assume an orthogonal rotator whose gravitational radiation is emitted at $2\nu$ without precession~\citep{ZimmermannSzedenits1979,JonesAndersson2002,KnispelAllen2008,LaskyMelatos2013}. In order to obtain $\dot{\nu}_{\rm gw} \approx \dot{\nu}$ (i.e. the measured spin-down rate) and hence $n_{\rm pl} \approx 5$, one requires $220 \leq (\epsilon/10^{-6})(I_{zz}/10^{45}~{\rm g\,cm}^{2})^{1/2}\leq 9.6\times10^{4}$ for the objects analyzed in this paper and defined in Section~\ref{Sec:Observations}. Even the lower extreme of the $\epsilon$ range stands well above the maximum values predicted by theory~\citep{GlampedakisGualtieri2018,MoralesHorowitz2022,Riles2023} and observational upper limits measured by the Laser Interferometer Gravitational Wave Observatory (LIGO), Virgo, and the Kamioka Gravitational Wave Detector (KAGRA)~\citep{Palomba2000,AbbottAbe2022,AbbottAbbott2022,AbbottAbe2022a,LuWette2023,Riles2023,TheLIGOScientificCollaborationtheVirgoCollaboration2025}. 

Gravitational radiation can also be emitted in the current quadrupole channel with $n_{\rm pl}\approx 7$~\citep{Thorne1980}. Typically stellar oscillations such as $r$-modes are invoked, albeit usually in accreting rather than isolated neutron stars, where specific excitation mechanisms have been identified~\citep{Andersson1998,OwenLindblom1998,AnderssonKokkotas2001,DongMelatos2024}. Less is known about the magnitudes of the current quadrupole moments in radio pulsars, and the ratio of the emission frequency to $\nu$ spans a wide range even for $r$-modes~\citep{CarideInta2019}. By equating the power emitted in current quadrupole gravitational radiation to the rotational energy loss rate, one obtains

\begin{equation}
    \dot{\nu}_{\rm gw} = -3.0\times10^{-14} \Bigg (\frac{\alpha}{10^{-3}}\Bigg)^{2}\Bigg (\frac{R_{\star}}{10~{\rm km}}\Bigg)^{6} \Bigg( \frac{\nu}{100~{\rm Hz}}\Bigg)^{7}~{\rm Hz\,s}^{-1}, \label{Eq_SecVI:sd_GWrmode}
\end{equation}

where $\alpha$ is the dimensionless $r$-mode amplitude~\citep{Thorne1980,OwenLindblom1998,Riles2023}. Terrestrial gravitational wave observations place constrains on $\alpha$, such as $\alpha \lesssim 10^{-1}$ for the young pulsar PSR J0537$-$6910~\citep{AbbottRmodes2021}, and $\alpha \lesssim 10^{-4}$ for millisecond pulsars in binary systems~\citep{CovasPapa2022}. $X$-ray observations imply $\alpha \lesssim 3\times10^{-9}$ for the millisecond pulsars PSR J1810$+$1744 and PSR J2241$-$5236~\citep{BoztepeGogus2020}. The constrains on $\alpha$ and equation~(\ref{Eq_SecVI:sd_GWrmode}) imply $ -3\times10^{-16}\lesssim \dot{\nu}_{\rm gw}/(1~{\rm Hz\;s}^{-1}) \lesssim -3\times10^{-31}$ for the analyzed P574 sample. That is, current quadrupole gravitational radiation reaction is unlikely to dominate magnetic dipole braking for any of the pulsars analyzed in this paper. The issue remains uncertain, however. Additional physics outside the scope of this paper may play a role~\citep{BlandfordRomani1988,AbolmasovBiryukov2024}, e.g. superfluid or radiative precession~\citep{Melatos2000, BarsukovTsygan2010,JonesDI2010}, magnetic state switching~\citep{BiryukovBeskin2012,Jones2012,StairsLyne2019} or internal torques~\citep{MelatosLink2014,GoglichidzeBarsukov2015}.  

\subsection{Single power-law mechanism with $\dot{K}_{\rm dim} \neq 0$} \label{subsecVI:single_pl_with_Kdot}

A single secular braking mechanism with $\dot{K}_{\rm dim} \neq 0$ is also consistent with the posterior $p(\mu_{\rm pl}, \sigma_{\rm pl}\vert D)$ calculated in Section~\ref{Sec:Analysis}. In this scenario, \citetalias{VargasMelatos2024} proved analytically, that the variance of the measured braking indices satisfies Equation~(\ref{Eq_SecI:Variance_n}) with $\dot{K}_{\rm dim} = (\dot{K}/K)(\nu/\dot{\nu}) \propto \tau_{\rm sd}/\tau_{K}$. 

The results in Section~\ref{Sec:Analysis} suggest, that $K$ increases ($\dot{K}_{\rm dim} >0$), if magnetic dipole braking or gravitational radiation reaction dominate, with $3 \lesssim n_{\rm pl} \lesssim 7$ and $n_{\rm pl} + \dot{K}_{\rm dim} \gtrsim 10$ (median value) for $40$ of the $68$ objects in Table~\ref{Tab:per-pulsar_inf}. If we restrict attention to the 10 objects whose secular anomaly dominates their stochastic anomaly, i.e.\ $|n_{\rm pl} + \dot{K}_{\rm dim} | \geq \chi s$, then we find $\dot{K}_{\rm dim} <0$ for PSR J1513$-$5908 and PSR J1833$-$0827, and $\dot{K}_{\rm dim} >0$ for the rest. 

Importantly, there is no way to distinguish observationally between $n_{\rm pl}$ and $\dot{K}_{\rm dim}$; they are degenerate because they always appear together inseparably in the sum $n_{\rm pl}+\dot{K}_{\rm dim}$ in the hierarchical Bayesian scheme (or indeed any other measurement scheme derived solely from a pulsar's spin-down behavior). For the sake of argument, however, it is interesting to follow through the implications of assuming $n_{\rm pl} \approx 3$, as magnetic dipole braking predicts, i.e. $\vert \dot{\nu}_{\rm gw} \vert \ll \vert \dot{\nu} \vert$; see Section~\ref{subsecVI:two_pl_with_no_Kdot}. Then $\dot{K}_{\rm dim} >0 $ is consistent with some published mechanisms of $K(t)$ evolution, including magnetic axis counter-alignment~\citep{Goldreich1970,AllenHorvath1997,Melatos2000}, and is inconsistent with others, including magnetic axis alignment and magnetic field dissipation~\citep{PonsVigano2012,GourgouliatosCumming2015}. We note also that $\dot{K}_{\rm dim} > 0$ is consistent with $p[(n_{\rm pl}+\dot{K}_{\rm dim})^{({\rm pop})} \vert D]$ (blue histogram in Fig.~\ref{fig_secIV:ppc}), with $82\%$ of the probability weight corresponding to $(n_{\rm pl}+\dot{K}_{\rm dim})^{({\rm pop})}>0$. The bias in favor of positive anomalous braking indices $n_{\rm meas}^{(m)}$ has been pointed out in observational studies published previously~\citep{LowerBailes2020, ParthasarathyJohnston2020,OnuchukwuLegahara2024}. The hieararchical Bayesian analysis in Sections~\ref{Sec:Analysis} and~\ref{Sec:Discussion} yields only one negative $(n_{\rm pl}+\dot{K}_{\rm dim})$-value, at $90\%$ confidence, for PSR J1833$-$0827.

The posterior predictive check performed in Fig.~\ref{fig_secIV:ppc} shows that $87\%$ of the probability weight lies outsides the interval $3 \leq (n_{\rm pl}+\dot{K}_{\rm dim})^{({\rm pop})} \leq 7$. At first glance, this represents additional corroborating evidence in favor of $\dot{K}_{\rm dim} \neq 0$. It is natural physically (see Section~\ref{Sec:Introduction}) to have $\vert n_{\rm pl}+\dot{K}_{\rm dim}\vert \gg 3$, for $\dot{K}_{\rm dim} \neq 0$, but it is less natural to have $\vert n_{\rm pl} \vert \gg 3$. However, the width of $p[(n_{\rm pl}+\dot{K}_{\rm dim})^{({\rm pop})} \vert D]$ cuts both ways. Although  $p[(n_{\rm pl}+\dot{K}_{\rm dim})^{({\rm pop})} \vert D]$ is relatively wide, it is fair to ask why it is not even wider. Many mechanisms of $K(t)$ evolution predict $| \dot{K}_{\rm dim} | \sim \tau_{\rm sd} / \tau_K \gg 1$, and hence $\vert n_{\rm pl}+\dot{K}_{\rm dim} \vert \gg n_{\rm pl}$. For example, magnetic axis (counter)alignment obeys $\tau_{K}/\tau_{\rm sd} \sim R_{\star}/R_{L} \approx 10^{-2}(\nu/10~{\rm Hz})$~\citep{Goldreich1970,Melatos2000}, which implies $-11 \lesssim n_{\rm pl}\pm \vert \dot{K}_{\rm dim} \vert \lesssim 230$, assuming $3 \lesssim n_{\rm pl} \lesssim 7$, for the pulsars analyzed in this paper. The inferred width of $p[(n_{\rm pl}+\dot{K}_{\rm dim})^{({\rm pop})} \vert D]$ in Fig.~\ref{fig_secIV:ppc} suggests instead that many pulsars in the sample obey $\tau_{K}/\tau_{\rm sd} \sim 1$ and hence $\vert n_{\rm pl}+\dot{K}_{\rm dim} \vert \gtrsim n_{\rm pl}$, if one has $\dot{K} \neq 0$. Arguably this favors slower evolutionary mechanisms, such as magnetic field dissipation,\footnote{On the other hand, it is hard to imagine that the Ohmic dissipation time-scale matches $\tau_{\rm sd}$ approximately in every pulsar, when the electrical conductivity and braking torque are unrelated physically.} which is predicted to obey $\tau_{K} \gtrsim 1~{\rm Myr}$~\citep{Romani1990,Bhattacharya2002,Faucher-GiguereKaspi2006,KielHurley2008}, and the variable electromagnetic torque in a force-free electron-positron outflow~\citep{ContopoulosSpitkovsky2006, Petri2021}. The former and latter mechanisms predict $\dot{K}_{\rm dim} <0$ and $\dot{K}_{\rm dim} > 0$, respectively. More data are needed to test these tentative conclusions further. 

\section{Conclusions} \label{Sec:Conclusions}

In this paper, we employ a hierarchical Bayesian scheme based on the successful Brownian model for anomalous braking indices (see Section~\ref{Sec:Methods}), to infer the posterior distribution $p(\mu_{\rm pl}, \sigma_{\rm pl} \vert D)$ of the torque law population hyperparameters, as well as the per-pulsar posterior distribution $p[(n_{\rm pl}+\dot{K}_{\rm dim})^{(m)},\chi^{(m)}\vert D]$, for a sample of 68 P574 pulsars~\citep{WeltevredeJohnston2010,ParthasarathyShannon2019,ParthasarathyJohnston2020,JohnstonSobey2021}. At the population level, the Bayesian scheme infers $\mu_{\rm pl} = 9.95^{+5.58}_{-5.26}$ and $\sigma_{\rm pl} = 10.89^{+5.14}_{-3.69}$, where the central value and error bars correspond to the median and $90\%$ credible intervals of the marginalized $p(\mu_{\rm pl}, \sigma_{\rm pl} \vert D)$ posterior. At the per-pulsar level, the inferred medians span the ranges $-13.86\leq (n_{\rm pl}+\dot{K}_{\rm dim})^{(m)} \leq 30.38$ and $-21.20 \leq \log_{10} \chi^{(m)}/(1~{\rm s}^{5/2}) \leq -18.26$. Table~\ref{Tab:per-pulsar_inf} summarizes the inferred medians and $90\%$ credible intervals for $\{(n_{\rm pl}+\dot{K}_{\rm dim})^{(m)},\chi^{(m)}\}_{1 \leq m \leq 68}$. Table~\ref{Tab:per-pulsar_inf} also quotes the stochastic braking index anomaly $\chi^{(m)} s^{(m)}$ [equation~(\ref{Eq_secII:nmeas_dist})]. We find that 58 out of 68 objects in the sample satisfy $\chi^{(m)}s^{(m)} \geq \vert (n_{\rm pl}+\dot{K}_{\rm dim})^{(m)}\vert$, with $\chi^{(m)}s^{(m)} \gg \vert (n_{\rm pl}+\dot{K}_{\rm dim})^{(m)}\vert$ in several instances. That is, the braking index anomaly is predominantly stochastic [scenario (ii) in Section~\ref{Sec:Introduction}] rather than secular [scenario (i) in Section~\ref{Sec:Introduction}]. In nine  objects in the top half of Table~\ref{Tab:per-pulsar_inf}, all of which satisfy $\chi s \leq 7$, the $(n_{\rm pl}+\dot{K}_{\rm dim})$-posteriors exclude zero at $90\%$ confidence. In one object, PSR J1833$-$0827, the inferred braking index satisfies $n_{\rm pl}+\dot{K}_{\rm dim} < 0$ at $90\%$ confidence. The inference results do not change markedly, when $13$ P574 objects with nonzero phase noise $\sigma_{\phi} \neq 0$ are added to the analysis, as demonstrated in Appendix~\ref{App:analysis_all_PSRs}.

In Section~\ref{Sec:Discussion}, we leverage the ability of the hierarchical Bayesian scheme to disentangle the stochastic and secular contributions of the inference results, to study interesting trends across the $68$ P574 pulsar sample. In terms of the location of pulsars in the $P$-$\dot{P}$ plane, we find that most pulsars whose measured braking indices are dominated by stochastic rather than secular anomalies, i.e. $\chi^{(m)} s^{(m)} \geq \vert (n_{\rm pl}+\dot{K}_{\rm dim})^{(m)}\vert$, are older objects which lie in the spin-down luminosity range $1 \lesssim \dot{E}/(10^{34}~{\rm erg\,s}^{-1}) \lesssim 10$. We find evidence that the inferred median $\chi$-value of a pulsar is related to its location in the $P$-$\dot{P}$ plane as $\chi \propto P^{-1.21\pm0.35}\dot{P}^{0.59\pm0.15}$ [see equation~(\ref{eq_secV:chiPPdot})] or equivalently $\chi \propto \nu^{0.038\pm0.34}\vert\dot{\nu}\vert^{0.59\pm0.15}$. In terms of correlations between first and second $\nu$-derivatives, we find that the stochastic-anomaly-free, median estimate $\ddot{\nu}_{\rm pl}^{(m)}= (n_{\rm pl}+\dot{K}_{\rm dim})^{(m)}[\dot{\nu}^{(m)}]^{2}[\nu^{(m)}]^{-1}$ strongly correlates with $\ddot{\nu}_{\rm dip}=\ddot{\nu}_{n_{\rm pl}=3, \dot{K}_{\rm dim}=0}=3\dot{\nu}^{2}\nu^{-1}$, $\vert \dot{\nu} \vert$, and $\tau_{\rm sd}=\nu/(2\dot{\nu})$. The correlations $\vert\ddot{\nu}_{\rm pl}\vert$-$\vert \dot{\nu}\vert$, $\vert\ddot{\nu}_{\rm pl}\vert$-$\tau_{\rm sd}$, and $\vert\ddot{\nu}_{\rm pl}\vert$-$\dot{\nu}_{\rm dip}$ span $5.6~{\rm dex}$ of $\vert \ddot{\nu}_{\rm pl} \vert$ values (see Fig.~\ref{fig_secV:correlations}). In contrast, the correlations are looser when using the second derivative $\vert \ddot{\nu}\vert$ measured by~\tempoDOS, which is contaminated by the stochastic anomaly, instead of $\vert \ddot{\nu}_{\rm pl}\vert$. Table~\ref{Tab:correlations} summarizes the Pearson correlation coefficients when using $\vert \ddot{\nu}\vert$ and $\vert \ddot{\nu}_{\rm pl}\vert$.

We discuss three plausible astrophysical interpretations for the inference results in Section~\ref{Sec:astro_int}, noting that many other interpretations are plausible too. The first scenario considers a single secular braking mechanism with $3 \lesssim n_{\rm pl} \lesssim 7$ and $\dot{K}_{\rm dim}=0$. Some of the inferred $n_{\rm pl}+\dot{K}_{\rm dim}=n_{\rm pl}$ values in Table~\ref{Tab:per-pulsar_inf} overlap with known physical mechanisms, e.g. a magnetic quadrupole superposed on a magnetic dipole yields $3 \leq n_{\rm pl} \leq 5$, as long as a suitable magnetic field formation mechanism exists, that deposits more energy into the quadrupole than the dipole. The second scenario considers two secular braking mechanisms with $3 \lesssim n_{\rm pl} \lesssim 7$ and $\dot{K}_{\rm dim}=0$. For example, $61$ objects in Table~\ref{Tab:per-pulsar_inf} return $(n_{\rm pl}+\dot{K}_{\rm dim})$-posteriors, which overlap with $n_{\rm pl} \approx 3$ and $n_{\rm pl} \approx 5$. However,  independent theoretical and observational bounds on the mass ellipticity $\epsilon$ make it unlikely that gravitational radiation reaction dominates electromagnetic radiation reaction in the P574 sample. To rigorously distinguish between one and two secular braking mechanisms, one needs to repeat the analysis in Section~\ref{Sec:Analysis} with $n_{\rm pl}^{(m)}$ drawn from a bimodal population-level distribution for a population size of $M \gtrsim 100$~\citepalias{VargasMelatos2025}. This is deferred to a forthcoming paper, when more data become available. The third scenario considers a single mechanism with $3 \lesssim n_{\rm pl} \lesssim 7$ and $\dot{K}_{\rm dim} \neq 0$. As tentative evidence for $\dot{K}_{\rm dim} \neq 0$, we find that $87\%$ of the probability weight of $p[(n_{\rm pl}+\dot{K}_{\rm dim})^{(\rm pop)}]$ lies outside $3\leq(n_{\rm pl}+\dot{K}_{\rm dim})^{(\rm pop)} \leq 7$, as do the median $(n_{\rm pl}+\dot{K}_{\rm dim})$ values of 62 out of the 68 objects in the P574 sample. Additionally, when considering the $10$ pulsars whose secular anomalies dominate their stochastic anomalies, i.e. $\vert n_{\rm pl}+\dot{K}_{\rm dim}\vert \geq \chi s$, we find that PSR J1513$-$5908 and PSR J1833$-$0827 support $\dot{K}_{\rm dim}<0$, while the rest support $\dot{K}_{\rm dim}>0$. Importantly, the inferred $n_{\rm pl}+\dot{K}_{\rm dim}$ values are consistent with $\tau_{K} \sim \tau_{\rm sd}$ and hence $\vert n_{\rm pl}+\dot{K}_{\rm dim}\vert \gtrsim n_{\rm pl}$ as opposed to $\tau_K \ll \tau_{\rm sd}$ and hence $|n_{\rm pl} + \dot{K}_{\rm dim}| \gg n_{\rm pl}$, which favors certain mechanisms referenced in Section~\ref{Sec:Introduction} over others. These three interpretations are not exhaustive, of course. Every conclusion drawn from the analysis in Section~\ref{Sec:Analysis} is conditional on the model (see Section~\ref{Sec:Methods}). We invite the reader to consider other plausible scenarios supported by the results in Table~\ref{Tab:per-pulsar_inf}.

A natural next step is to analyze a larger sample. This may reveal interesting $P$-$\dot{P}$ trends beyond those identified in Section~\ref{Sec:Discussion}. It will also produce more accurate inference results. \citetalias{VargasMelatos2025} demonstrated through Monte Carlo simulations, that the percentage of $(n_{\rm pl}+\dot{K}_{\rm dim})$-posteriors which include the injected $(n_{\rm pl}+\dot{K}_{\rm dim})^{(m)}$ value within the $90\%$ credible intervals improves from $\approx 66\%$ for $M=50$ to $\approx 87\%$ for $M=100$. The agreement between $n_{\rm pl}+\dot{K}_{\rm dim}$ inferred in this paper (second column in Table~\ref{Tab:per-pulsar_inf}) and braking indices measured previously (last column in Table~\ref{Tab:per-pulsar_inf}) for objects with $\vert n_{\rm pl}+\dot{K}_{\rm dim}\vert \gtrsim \chi s$ confirms that the hierarchical Bayesian scheme complements rather than replaces traditional timing software. It can be employed in future timing campaigns to measure secular $(n_{\rm pl}+\dot{K}_{\rm dim})$ and stochastic ($\chi s$) braking index anomalies in tandem with traditional timing methods.

\section*{Acknowledgments} 

This research was supported by the Australian Research Council Centre of Excellence for Gravitational Wave Discovery (OzGrav), grant number CE170100004. The numerical calculations were performed on the OzSTAR supercomputer facility at Swinburne University of Technology. The OzSTAR program receives funding in part from the Astronomy National Collaborative Research Infrastructure Strategy (NCRIS) allocation provided by the Australian Government. Murriyang, CSIRO’s Parkes radio telescope is part of the Australia Telescope National Facility (\href{https://ror.org/05qajvd42}{https://ror.org/05qajvd42}) which is funded by the Australian Government for operation as a National Facility managed by CSIRO. We acknowledge the Wiradjuri people as the traditional owners of the Observatory site. MEL is supported by the ARC Discovery Early Career Research Award DE250100508. 

\section*{Data availability}

The raw Parkes data is available to download via the CSIRO Data Access Portal (\href{https://data.csiro.au/}{https://data.csiro.au/}). Other data products are available upon reasonable request to the corresponding author.


\twocolumn

\bibliographystyle{mnras}
\bibliography{ADSABS_bib, main_non_ads_bib}


\appendix

\section{Calibration of the per-pulsar $S_{\rm \MakeLowercase{meas}}^{(\MakeLowercase{m})}$ likelihood} \label{App:S_m_calibration}

The per-pulsar $S_{\rm meas}^{(m)}$-likelihood [equation~(\ref{Eq_secII:Smeas_dist})] is a fundamental input into the hierarchical Bayesian scheme described in Section~\ref{subsecII:Bayesframework}. \citetalias{VargasMelatos2025} showed through analytic calculations and Monte Carlo simulations involving synthetic data, that $\log S_{\rm meas}^{(m)}$ follows approximately a normal distribution with mean $\mu_{S,{\rm BM}}[\chi^{(m)}]$ and standard deviation $\sigma_{S,{\rm BM}}$. In this appendix, we briefly show how to calculate $\mu_{S,{\rm BM}}[\chi^{(m)}]$ and $\sigma_{S,{\rm BM}}$. For a thorough discussion, the reader is referred to Appendix B of \citetalias{VargasMelatos2025}.

The standard deviation $[S^{(m)}_{\rm meas}]^{2}$ [i.e. equation~(\ref{eq_subsecIII:S_meas_m})] of the timing residuals ${\cal R} [t^{(m)}_{i}]$ obtained by fitting a polynomial ephemeris [up to $\ddot{\nu}^{(m)}$; see Section~\ref{subsecIII:measurements_D^m}] is a standard indicator of timing noise strength~\citep{CordesHelfand1980, HobbsLyne2010, ShannonCordes2010,AntonelliBasu2023}. For the Brownian model, every realization of the stochastic process $d{\bf B}(t)$ in equation~(\ref{Eq_secII:dX}), at a fixed $\chi^{(m)}=\sigma^{(m)}[\gamma_{\ddot{\nu}}^{(m)}]^{-1}$ value, yields a statistically equivalent set of TOAs. The ensemble's variance $\langle [S^{(m)}_{\rm meas}]^{2} \rangle$, in the astrophysically relevant regime $\gamma_{\nu}^{(m)}\sim\gamma_{\dot{\nu}}^{(m)} \ll [T_{\rm obs}^{(m)}]^{-1} \ll \gamma_{\ddot{\nu}}^{(m)}$, is given by

\begin{equation}
    \langle [S^{(m)}_{\rm meas}]^{2} \rangle = \frac{[\chi^{(m)}]^{2}[T_{\rm obs}^{(m)}]^{5}}{160\pi^{6}[\nu^{(m)}]^{2}}. \label{eq_AppA: var_S_meas_chi}
\end{equation}

In general, equation~(\ref{eq_AppA: var_S_meas_chi}) needs to be multiplied by a dimensionless constant, $C_{0}$, to match the ensemble variance measured by \tempoDOS, $\langle [S^{(m)}_{\rm meas}]^{2} \rangle_{\rm T2} \approx C_{0}\langle [S^{(m)}_{\rm meas}]^{2} \rangle$. Similarly, we assume that the mean $\langle S^{(m)}_{\rm meas} \rangle_{\rm T2}$ is given by $\langle S^{(m)}_{\rm meas} \rangle_{\rm T2} \approx C_{1}\sqrt{\langle [S^{(m)}_{\rm meas}]^{2} \rangle}$. The constants $C_{0}$ and $C_{1}$ are empirically determined by calibrating the output of equation~(\ref{eq_AppA: var_S_meas_chi}) with the measured $\langle [S^{(m)}_{\rm meas}]^{2} \rangle_{\rm T2}$ and $\langle S^{(m)}_{\rm meas} \rangle_{\rm T2}$ for an ensemble of synthetic TOA realizations generated using a selection of typical pulsar parameters, e.g. $\chi^{(m)}, {\bf X}^{(m)}$, and $T_{\rm obs}^{(m)}$. On average, we find $C_{0} \approx 6.3\times10^{-2}$ and $C_{1} \approx 4.5\times10^{-1}$. The calibration step should be repeated when using a different timing software package and/or polynomial fit.  

Once calibrated, equation~(\ref{eq_AppA: var_S_meas_chi}) determines the mean and variance of the per-pulsar $S_{\rm meas}^{(m)}$-likelihood as

 \begin{equation}
     \mu_{S \rm, BM}[\chi^{(m)}] = \ln\Bigg\{C_{1}\sqrt{\langle S_{\rm meas}^{2} \rangle[\chi^{(m)}]}\Bigg\}-\frac{1}{2}\ln\Bigg(1+\frac{C_{0}}{C_{1}^{2}}\Bigg), \label{eq:mu_SBM}
 \end{equation}

and 

\begin{equation}
    \sigma^{2}_{S \rm,BM} = \ln\Bigg(1+\frac{C_{0}}{C_{1}^{2}}\Bigg).\label{eq:sigma_SBM}
\end{equation}

\section{Uncertainties in $\Delta \ddot{\nu}$ from~\tempoDOS~and~\temponest}
\label{App:uncF2}
The uncertainties for $\Delta \ddot{\nu}$ returned by~\tempoDOS~differ sometimes by a factor $\lesssim 3$ from the more accurate uncertainties returned by~\temponest, which include spin noise and dispersion measure variations~\citep{LentatiAlexander2014,ParthasarathyJohnston2020}. Despite this, we stick with~\tempoDOS~for the analysis in this paper, because the condition $\Delta n_{\rm meas} \propto \Delta\ddot{\nu} \lesssim \chi s$ holds for all 68 pulsars. That is, the variance of the measured braking indices [see Eq.~(\ref{Eq_secII:nmeas_dist})] is dominated by spin noise, through the stochastic anomaly $\chi s$, and not the uncertainties in the rotational parameters $\Delta \nu, \Delta \dot{\nu},$ and $\Delta \ddot{\nu}$. Specifically, and by the way of verification, we compare with the results of~\cite{ParthasarathyJohnston2020} and find $\Delta n_{\rm meas} \ll \chi s$ for 62 pulsars and $\Delta n_{\rm meas} \sim \chi s$ for six pulsars (where~\tempoDOS~underestimates $\Delta \ddot{\nu}$ somewhat). Despite the underestimate, however, Table~\ref{Tab:per-pulsar_inf} confirms that even for these six pulsars the estimates of $n_{\rm pl}+\dot{K}_{\rm dim}$ are consistent with those of $n$ obtained using~\temponest~\citep{ParthasarathyJohnston2020}. For instance, for PSR J1513$-$5908 we find $n_{\rm pl}+\dot{K}_{\rm dim}=2.83^{+0.01}_{-0.01}$ and the~\temponest~estimate is $n=2.82^{+0.06}_{-0.06}$~\citep{ParthasarathyJohnston2020}. If the need arises, say when analyzing a pulsar population with $\Delta n_{\rm meas} \gg \chi s$ for most objects, one can use other timing software to fit the TOAs, e.g.~\temponest, by appropriately repeating the calibration of the $S_{\rm meas}^{(m)}$-likelihood parameters $C_{0}$ and $C_{1}$ as discussed in Appendix~\ref{App:S_m_calibration}. However, repeating the calibration with another timing software may come at the expense of significantly higher computational cost.

\section{Including pulsars with $\sigma_{\phi} \neq 0$} \label{App:analysis_all_PSRs}

\begin{figure}
\flushleft
 \includegraphics[width=\columnwidth]{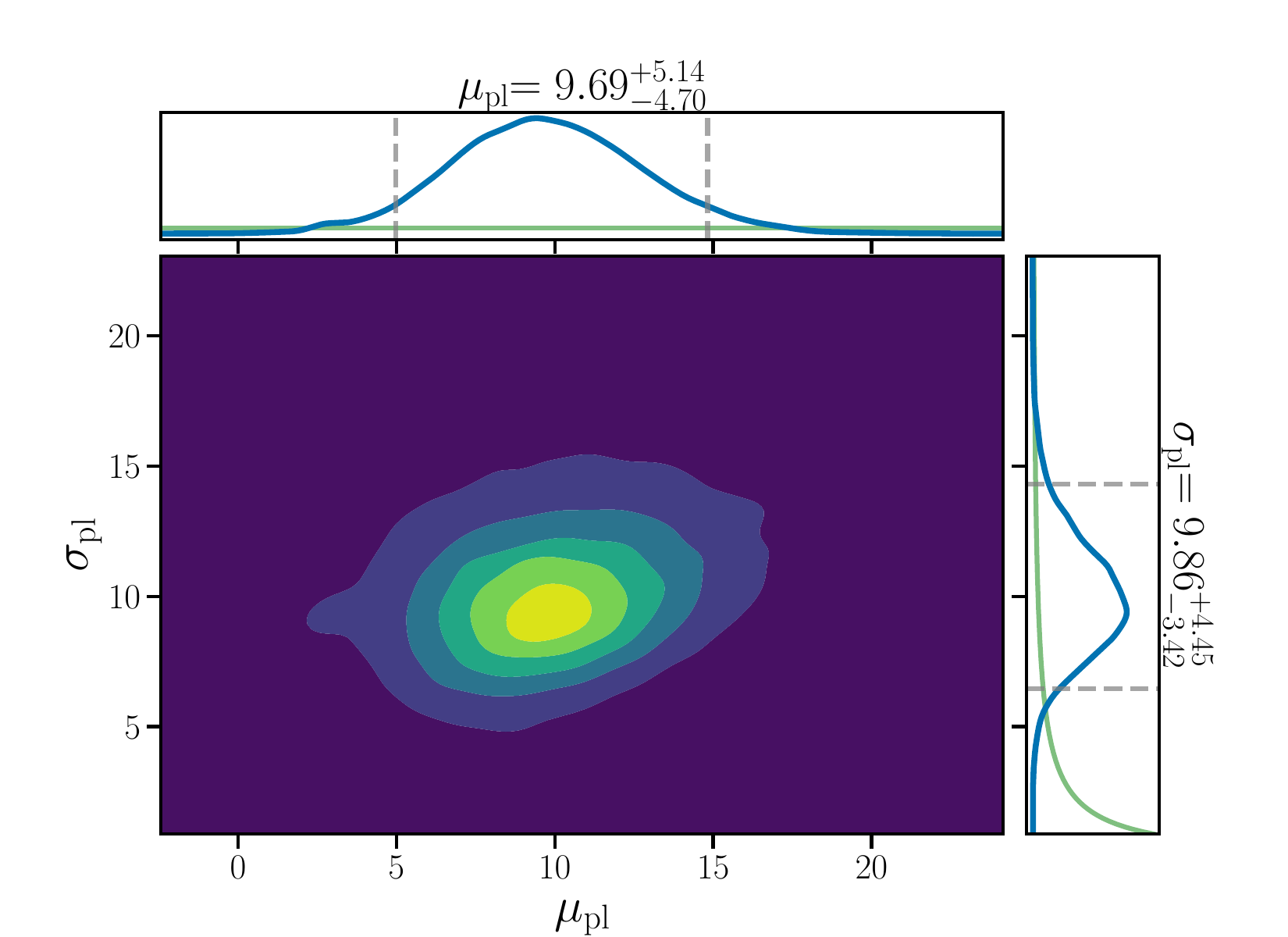}
 \caption{Posterior distribution $p(\mu_{\rm pl}, \sigma_{\rm pl} \vert D)$ (central panel) for the population-level hyperparameters $\mu_{\rm pl}$ and $\sigma_{\rm pl}$ for a subsample of P574 pulsars that includes the 68 objects analyzed in Sections~\ref{Sec:Analysis} and~\ref{Sec:Discussion} as well as another $13$ pulsars with nonzero phase noise $\sigma_\phi \neq 0$ ($M=81$; see Table~\ref{AppB:TableD^m}). The credible intervals and priors are colour-coded as in Fig.~\ref{fig_secIV:corner_plot}.}
\label{fig_AppB:non_glitch_corner_plot}
\end{figure}

In this section, we make a first attempt at generalizing the analysis in Section~\ref{Sec:Analysis} to include the $13$ P574 objects, whose timing residuals are qualitatively consistent with noise in the rotational phase $\sigma_{\phi} \neq 0$. That is,  we analyze a P574 subsample with $M=68+13=81$, which excludes objects that glitched during the observation campaign. The $13$ additional objects and their $D^{(m)}$ entries are summarized in the lower half of Table~\ref{AppB:TableD^m}.

Fig.~\ref{fig_AppB:non_glitch_corner_plot} displays the two-dimensional posterior for the hyperparameters, $p(\mu_{\rm pl}, \sigma_{\rm pl}\vert D)$, presented in the same fashion as Fig.~\ref{fig_secIV:corner_plot}. The revised parameter estimates $\mu_{\rm pl} = 9.69^{+5.14}_{-4.70}$ and $\sigma_{\rm pl}=9.86^{+4.45}_{-3.42}$ overlap with the $M=68$ estimates $\mu_{\rm pl} = 9.95^{+5.58}_{-5.26}$ and $\sigma_{\rm pl} = 10.89^{+5.14}_{-3.69}$. As in Section~\ref{Sec:Analysis}, the error bars on $\mu_{\rm pl}$ exclude the combination $n_{\rm pl}=3$ and $\dot{K}=0$ at $90\%$ confidence. 

At a per-pulsar level, the central values of the one-dimensional posteriors for $\theta^{(m)}=\{(n_{\rm pl}+\dot{K}_{\rm dim})^{(m)}, \chi^{(m)}\}$ span the ranges $-13.70 \leq (n_{\rm pl}+\dot{K}_{\rm dim})^{(m)} \leq 28.70$ and $-21.54 \leq \log_{10} \chi^{(m)}/(1~{\rm s}^{5/2}) \leq -18.26$. For the objects shared with the $M=68$ sample, the inferred $\theta^{(m)}$-values overlap considerably with those listed in Table~\ref{Tab:per-pulsar_inf}. For the excluded $13$ objects, the inferred $\theta^{(m)}$-values span the ranges $6.49 \leq (n_{\rm pl}+\dot{K}_{\rm dim})^{(m)} \leq 10.79$ and $-21.54 \leq \log_{10} \chi^{(m)}/(1~{\rm s}^{5/2}) \leq -19.71$. The stochastic anomalies span $4 \leq \chi^{(m)}s^{(m)} \leq 1511$. Interestingly, out of these $13$ pulsars, only PSR J1838$-$0549 with $n_{\rm pl}+\dot{K}_{\rm dim}=9.96^{+8.15}_{-8.15}$ excludes zero at $90\%$ confidence and satisfies $\chi s \leq n_{\rm pl}+\dot{K}_{\rm dim}$, i.e. its secular anomaly dominates its stochastic anomaly. For this object, the reported $95\%$ credible interval for $n$ is $10 \leq n \leq 15$~\citep{ParthasarathyShannon2019}, which is bracketed by the inferred $n_{\rm pl}+\dot{K}_{\rm dim}=9.96^{+8.15}_{-8.15}$. Only one other object, PSR J1452$-$5851, returns $\chi s \leq n_{\rm pl}+\dot{K}_{\rm dim}$, with $n_{\rm pl}+\dot{K}_{\rm dim}=6.49^{+6.72}_{-7.51}$. 

The alignment between the results in this appendix and Sections~\ref{Sec:Analysis} and~\ref{Sec:Discussion} does not alter the astrophysical interpretation in Section~\ref{Sec:astro_int}.

\section{$D^{(\MakeLowercase{m})}$ for $M=81$ P574 pulsars}
\label{subsec_AppB:D^m_for_68_psrs}

 In this appendix, the data for the $M=81$ P574 objects analyzed in Sections~\ref{Sec:Analysis} and~\ref{Sec:Discussion} and Appendix~\ref{App:analysis_all_PSRs} are presented in Table~\ref{AppB:TableD^m} for the sake of reproducibility~\citep{ParthasarathyShannon2019,ParthasarathyJohnston2020}. Every row in the table quotes $D^{(m)} = \{n_{\rm meas}, \Delta n_{\rm meas}, S_{\rm meas},\Delta S_{\rm meas}\}$ for one of the $M$ objects. The data $D = \{ D^{(m)} \}_{1\leq m \leq 81}$ are fed into the hierarchical Bayesian scheme, as described in Sections~\ref{Sec:Methods} and~\ref{Sec:Observations}.

\begin{table*}
\begin{center}
\caption{ Braking index data and auxiliary timing data collected by the Murriyang P574 observing campaign~\citep{ParthasarathyShannon2019,ParthasarathyJohnston2020} and analyzed by the hierarchical Bayesian scheme in this paper. Every row contains the raw data $D^{(m)}$ for one pulsar. The name of the pulsar appears in column 1. Columns 2 and 3 quote the braking index measured by~\tempoDOS~and its formal uncertainty respectively. Columns 4 and 5 quote the root-mean-square of the timing residuals and its associated uncertainty (see Section~\ref{subsecIII:measurements_D^m}) respectively. The top segment of the table above the horizontal line summarizes the subsample with $M=68$ studied in Section~\ref{Sec:Analysis} and \ref{Sec:Discussion}. The lower segment features $M=13$ additional objects with nonzero phase noise $\sigma_\phi\neq 0$, which are combined with the objects in the upper segment and analyzed together in Appendix~\ref{App:analysis_all_PSRs}.}
\label{AppB:TableD^m}
\renewcommand{\arraystretch}{1.65}  
\begin{tabular*}{\textwidth}{@{\extracolsep{\fill}}lcccc}
\hline
PSR & $n_{\rm meas}$ & $\Delta n_{\rm meas}$ & $S_{\rm meas}/(1~{\rm s})$ & $\Delta S_{\rm meas}/(1~{\rm s})$ \\  
\hline
J0543$+$2329 & $7.69\times10^{-1}$ & $3.46\times10^{-2}$ & $9.85\times10^{-4}$ & $7.63\times10^{-8}$ \\
 J0745$-$5353 & $2.12\times10^{2}$ & $8.99\times10^{-1}$ & $5.33\times10^{-4}$ & $9.87\times10^{-9}$ \\
 J0820$-$3826 & $-2.61\times10^{1}$ & $8.61\times10^{-1}$ & $2.41\times10^{-3}$ & $1.26\times10^{-7}$ \\
 J0834$-$4159 & $-9.93\times10^{0}$ & $1.54\times10^{-1}$ & $7.11\times10^{-4}$ & $5.43\times10^{-8}$ \\
 J0857$-$4424 & $2.89\times10^{3}$ & $4.82\times10^{-2}$ & $2.6\times10^{-2}$ & $1.06\times10^{-6}$ \\
 J0905$-$5127 & $-8.87\times10^{1}$ & $1.55\times10^{-2}$ & $3.18\times10^{-2}$ & $6.55\times10^{-7}$ \\
 J0954$-$5430 & $1.48\times10^{1}$ & $2.53\times10^{-3}$ & $3.13\times10^{-3}$ & $1.14\times10^{-7}$ \\
 J1043$-$6116 & $5.97\times10^{1}$ & $5.33\times10^{-2}$ & $1.89\times10^{-3}$ & $2.51\times10^{-8}$ \\
 J1115$-$6052 & $-1.1\times10^{1}$ & $1.67\times10^{-1}$ & $3.21\times10^{-3}$ & $9.2\times10^{-8}$ \\
 J1123$-$6259 & $5.66\times10^{2}$ & $4.97\times10^{-1}$ & $2.11\times10^{-2}$ & $1.06\times10^{-6}$ \\
 J1156$-$5707 & $-4.21\times10^{2}$ & $2.46\times10^{-2}$ & $8.03\times10^{-2}$ & $4.52\times10^{-6}$ \\
 J1224$-$6407 & $8.05\times10^{-1}$ & $1.48\times10^{-1}$ & $1.01\times10^{-3}$ & $1.82\times10^{-8}$ \\
 J1305$-$6203 & $1.87\times10^{-1}$ & $8.54\times10^{-2}$ & $1.92\times10^{-3}$ & $2.48\times10^{-7}$ \\
 J1349$-$6130 & $-2.66\times10^{2}$ & $3.01\times10^{-1}$ & $8.32\times10^{-3}$ & $2.52\times10^{-7}$ \\
 J1412$-$6145 & $1.27\times10^{1}$ & $1.78\times10^{-3}$ & $4.1\times10^{-2}$ & $4.29\times10^{-6}$ \\
 J1509$-$5850 & $1.12\times10^{1}$ & $4.48\times10^{-3}$ & $1.94\times10^{-3}$ & $1.37\times10^{-7}$ \\
 J1512$-$5759 & $-2.34\times10^{1}$ & $1.45\times10^{-2}$ & $6.24\times10^{-3}$ & $5.64\times10^{-8}$ \\
 J1513$-$5908 & $2.83\times10^{0}$ & $8.52\times10^{-7}$ & $1.28\times10^{-1}$ & $1.33\times10^{-5}$ \\
 J1514$-$5925 & $1.18\times10^{2}$ & $3.13\times10^{0}$ & $2.44\times10^{-3}$ & $2.11\times10^{-7}$ \\
 J1515$-$5720 & $9.76\times10^{1}$ & $3.04\times10^{-1}$ & $3.83\times10^{-3}$ & $1.36\times10^{-7}$ \\
 J1524$-$5706 & $4.19\times10^{0}$ & $8.53\times10^{-4}$ & $4.34\times10^{-3}$ & $5.12\times10^{-7}$ \\
 J1530$-$5327 & $-6.56\times10^{1}$ & $4.24\times10^{-1}$ & $9.06\times10^{-4}$ & $2.89\times10^{-8}$ \\
 J1531$-$5610 & $4.47\times10^{1}$ & $2.11\times10^{-3}$ & $1.34\times10^{-2}$ & $2.09\times10^{-7}$ \\
 J1539$-$5626 & $5.33\times10^{2}$ & $1.56\times10^{-1}$ & $1.44\times10^{-2}$ & $2.03\times10^{-7}$ \\
 J1543$-$5459 & $3.59\times10^{1}$ & $1.5\times10^{-2}$ & $4.84\times10^{-2}$ & $3.73\times10^{-6}$ \\
 J1548$-$5607 & $3.02\times10^{1}$ & $2.75\times10^{-2}$ & $6.92\times10^{-3}$ & $1.96\times10^{-7}$ \\
 J1549$-$4848 & $1.18\times10^{2}$ & $4.66\times10^{-2}$ & $6.29\times10^{-3}$ & $2.64\times10^{-7}$ \\
 J1551$-$5310 & $6.6\times10^{1}$ & $2.6\times10^{-2}$ & $1.16\times10^{-1}$ & $3.97\times10^{-5}$ \\
 J1600$-$5751 & $4.58\times10^{2}$ & $8.83\times10^{-1}$ & $4.13\times10^{-3}$ & $1.18\times10^{-7}$ \\
 J1601$-$5335 & $1.39\times10^{1}$ & $3.85\times10^{-2}$ & $8.79\times10^{-3}$ & $1.06\times10^{-6}$ \\
 J1611$-$5209 & $-1.1\times10^{2}$ & $3.43\times10^{-2}$ & $6.04\times10^{-3}$ & $5.22\times10^{-8}$ \\
 J1632$-$4757 & $8.09\times10^{1}$ & $6.73\times10^{-1}$ & $6.67\times10^{-3}$ & $1.42\times10^{-6}$ \\
 J1637$-$4553 & $1.07\times10^{2}$ & $6.33\times10^{-2}$ & $5.74\times10^{-3}$ & $8.27\times10^{-8}$ \\
 J1637$-$4642 & $3.57\times10^{1}$ & $3.29\times10^{-3}$ & $1.78\times10^{-2}$ & $2.14\times10^{-6}$ \\
 J1638$-$4417 & $6.61\times10^{1}$ & $1.25\times10^{0}$ & $2.09\times10^{-3}$ & $1.15\times10^{-7}$ \\
 J1640$-$4715 & $7.37\times10^{1}$ & $9.84\times10^{-2}$ & $4.73\times10^{-2}$ & $7.54\times10^{-6}$ \\
\hline
\end{tabular*}
\end{center}
\end{table*}

\begin{table*}
\begin{center}
\contcaption{}
\renewcommand{\arraystretch}{1.65}  
\begin{tabular*}{\textwidth}{@{\extracolsep{\fill}}lcccc}
\hline
PSR & $n_{\rm meas}$ & $\Delta n_{\rm meas}$ & $S_{\rm meas}/(1~{\rm s})$ & $\Delta S_{\rm meas}/(1~{\rm s})$ \\  
\hline
  J1643$-$4505 & $1.54\times10^{1}$ & $2.03\times10^{-2}$ & $4.79\times10^{-3}$ & $3.36\times10^{-7}$ \\
 J1648$-$4611 & $2.68\times10^{1}$ & $2.55\times10^{-2}$ & $1.32\times10^{-2}$ & $1.76\times10^{-6}$ \\
 J1649$-$4653 & $4.81\times10^{1}$ & $8.81\times10^{-2}$ & $9.12\times10^{-2}$ & $1.56\times10^{-5}$ \\
 J1702$-$4306 & $3.99\times10^{1}$ & $1.82\times10^{-1}$ & $2.93\times10^{-3}$ & $1.98\times10^{-7}$ \\
 J1722$-$3712 & $-2.15\times10^{2}$ & $1.06\times10^{-1}$ & $3.93\times10^{-2}$ & $2.1\times10^{-6}$ \\
 J1723$-$3659 & $1.89\times10^{1}$ & $8.65\times10^{-2}$ & $2.39\times10^{-2}$ & $8.07\times10^{-7}$ \\
 J1733$-$3716 & $2.31\times10^{1}$ & $6.63\times10^{-2}$ & $2.74\times10^{-3}$ & $1.07\times10^{-7}$ \\
 J1735$-$3258 & $-3.05\times10^{1}$ & $1.46\times10^{0}$ & $3.07\times10^{-2}$ & $1.68\times10^{-5}$ \\
 J1738$-$2955 & $-6.62\times10^{1}$ & $5.39\times10^{-2}$ & $2.16\times10^{-2}$ & $3.24\times10^{-6}$ \\
 J1739$-$2903 & $1.68\times10^{1}$ & $9.48\times10^{-2}$ & $7.82\times10^{-4}$ & $1.35\times10^{-8}$ \\
 J1739$-$3023 & $2.95\times10^{0}$ & $1.78\times10^{-2}$ & $6.1\times10^{-3}$ & $2.61\times10^{-7}$ \\
 J1745$-$3040 & $-6.35\times10^{0}$ & $4.73\times10^{-2}$ & $2.63\times10^{-3}$ & $5.02\times10^{-8}$ \\
 J1801$-$2154 & $2.44\times10^{2}$ & $1.3\times10^{0}$ & $5.03\times10^{-3}$ & $7.54\times10^{-7}$ \\
 J1806$-$2125 & $-4.04\times10^{0}$ & $1.44\times10^{-2}$ & $5.11\times10^{-2}$ & $1.7\times10^{-5}$ \\
 J1809$-$1917 & $2.43\times10^{1}$ & $1.37\times10^{-3}$ & $1.69\times10^{-2}$ & $5.67\times10^{-7}$ \\
 J1815$-$1738 & $9.56\times10^{0}$ & $9.49\times10^{-3}$ & $3.54\times10^{-3}$ & $3.25\times10^{-7}$ \\
 J1820$-$1529 & $6.97\times10^{1}$ & $2.38\times10^{-1}$ & $3.13\times10^{-2}$ & $1.07\times10^{-5}$ \\
 J1824$-$1945 & $1.25\times10^{2}$ & $8.41\times10^{-3}$ & $4.34\times10^{-2}$ & $2.74\times10^{-7}$ \\
 J1825$-$1446 & $3.9\times10^{1}$ & $3.51\times10^{-2}$ & $1.39\times10^{-2}$ & $7.86\times10^{-7}$ \\
 J1828$-$1101 & $1.26\times10^{1}$ & $2.14\times10^{-2}$ & $7.0\times10^{-2}$ & $2.08\times10^{-5}$ \\
 J1832$-$0827 & $2.0\times10^{-2}$ & $1.91\times10^{-2}$ & $2.93\times10^{-3}$ & $1.39\times10^{-7}$ \\
 J1833$-$0827 & $-1.46\times10^{1}$ & $9.19\times10^{-4}$ & $2.02\times10^{-3}$ & $2.62\times10^{-8}$ \\
 J1834$-$0731 & $2.94\times10^{0}$ & $3.24\times10^{-1}$ & $5.17\times10^{-3}$ & $1.45\times10^{-6}$ \\
 J1835$-$1106 & $-5.08\times10^{1}$ & $1.32\times10^{-2}$ & $5.61\times10^{-2}$ & $2.62\times10^{-6}$ \\
 J1837$-$0559 & $-1.05\times10^{2}$ & $1.71\times10^{0}$ & $6.27\times10^{-3}$ & $8.07\times10^{-7}$ \\
 J1838$-$0453 & $-1.03\times10^{2}$ & $2.7\times10^{-2}$ & $1.12\times10^{-1}$ & $1.89\times10^{-5}$ \\
 J1839$-$0321 & $1.04\times10^{2}$ & $7.21\times10^{-1}$ & $2.52\times10^{-3}$ & $3.11\times10^{-7}$ \\
 J1839$-$0905 & $3.21\times10^{2}$ & $1.31\times10^{0}$ & $1.2\times10^{-3}$ & $1.12\times10^{-7}$ \\
 J1842$-$0905 & $-8.45\times10^{1}$ & $1.53\times10^{-1}$ & $8.69\times10^{-3}$ & $3.11\times10^{-7}$ \\
 J1843$-$0702 & $3.27\times10^{2}$ & $1.58\times10^{0}$ & $4.38\times10^{-3}$ & $2.18\times10^{-7}$ \\
 J1844$-$0538 & $4.79\times10^{1}$ & $1.12\times10^{-1}$ & $2.79\times10^{-3}$ & $1.06\times10^{-7}$ \\
 J1853$-$0004 & $2.41\times10^{1}$ & $2.37\times10^{-2}$ & $5.0\times10^{-2}$ & $9.47\times10^{-7}$ \\
\hline
J1016$-$5819 & $4.36\times10^{1}$ & $2.01\times10^{0}$ & $1.92\times10^{-4}$ & $5.32\times10^{-9}$ \\
 J1020$-$6026 & $1.63\times10^{1}$ & $1.87\times10^{0}$ & $9.82\times10^{-4}$ & $2.87\times10^{-7}$ \\
 J1216$-$6223 & $1.16\times10^{1}$ & $1.09\times10^{0}$ & $7.66\times10^{-4}$ & $1.34\times10^{-7}$ \\
 J1452$-$5851 & $5.95\times10^{0}$ & $1.29\times10^{-1}$ & $5.47\times10^{-4}$ & $9.28\times10^{-8}$ \\
 J1453$-$6413 & $9.29\times10^{1}$ & $1.16\times10^{-1}$ & $1.06\times10^{-3}$ & $7.5\times10^{-9}$ \\
 J1538$-$5551 & $3.8\times10^{1}$ & $1.06\times10^{0}$ & $4.91\times10^{-4}$ & $3.61\times10^{-8}$ \\
 J1650$-$4921 & $4.77\times10^{1}$ & $1.2\times10^{0}$ & $2.09\times10^{-4}$ & $7.98\times10^{-9}$ \\
 \hline
\end{tabular*}
\end{center}
\end{table*}

\begin{table*}
\begin{center}
\contcaption{}
\renewcommand{\arraystretch}{1.65}  
\begin{tabular*}{\textwidth}{@{\extracolsep{\fill}}lcccc}
\hline
PSR & $n_{\rm meas}$ & $\Delta n_{\rm meas}$ & $S_{\rm meas}/(1~{\rm s})$ & $\Delta S_{\rm meas}/(1~{\rm s})$ \\  
\hline
J1828$-$1057 & $8.02\times10^{0}$ & $5.48\times10^{-1}$ & $1.21\times10^{-3}$ & $3.74\times10^{-7}$ \\
 J1835$-$0944 & $1.11\times10^{2}$ & $8.28\times10^{0}$ & $3.88\times10^{-4}$ & $4.26\times10^{-8}$ \\
 J1838$-$0549 & $1.0\times10^{1}$ & $1.14\times10^{-1}$ & $7.22\times10^{-4}$ & $1.51\times10^{-7}$ \\
 J1843$-$0355 & $-4.34\times10^{1}$ & $5.23\times10^{1}$ & $1.21\times10^{-3}$ & $3.32\times10^{-7}$ \\
 J1845$-$0743 & $-1.88\times10^{1}$ & $1.57\times10^{0}$ & $5.26\times10^{-5}$ & $2.91\times10^{-10}$ \\
 J1853$+$0011 & $5.44\times10^{0}$ & $3.99\times10^{0}$ & $6.79\times10^{-4}$ & $2.12\times10^{-7}$ \\
 \hline
\end{tabular*}
\end{center}
\end{table*}

\bsp	
\label{lastpage}
\end{document}